\documentclass[journal]{IEEEtran}
\usepackage{graphicx}
\usepackage{cite}
\usepackage{amsmath,amssymb,amsfonts,amssymb}
\usepackage{graphicx}
\usepackage{textcomp}
\usepackage{xcolor}
\usepackage{mathrsfs}
\usepackage{caption}
\usepackage{array}
\allowdisplaybreaks  
\usepackage{graphicx}
\usepackage{verbatim}
\usepackage{enumitem}
\usepackage{booktabs}
\usepackage{tabularx}
\usepackage{bm}
\usepackage{color}
\usepackage{hyperref}
\hypersetup{hidelinks}

\usepackage{cite}
\usepackage{amsmath}
\usepackage{psfrag}
\usepackage{amsfonts}
\usepackage{graphicx}
\usepackage{multirow}
\usepackage{float}
\usepackage{color} 
\usepackage{stfloats}
\usepackage{balance}
\usepackage{bm}
\usepackage{cuted}
\usepackage{dutchcal}
\usepackage[none]{hyphenat}
\usepackage[lined,ruled,commentsnumbered]{algorithm2e}
\usepackage{diagbox}
\usepackage{array}
\usepackage{booktabs}
\usepackage{url}
\usepackage{subfigure}
\usepackage{caption}
\allowdisplaybreaks[4]
\renewcommand{\arraystretch}{1.3}
\newcommand\scalemath[2]{\scalebox{#1}{\mbox{\ensuremath{\displaystyle #2}}}}

\newtheorem{theorem}{Theorem}
\newtheorem{lemma}{Lemma}
\newtheorem{definition}{Definition}

\newtheorem{remark}{Remark}

\newtheorem{corollary}{Corollary}

\newcommand{\x}{\mathbf{x}}
\newcommand{\y}{\mathbf{y}}
\newcommand{\I}{\mathbf{I}}

\newcommand{\p}{\mathbf{p}}

\renewcommand{\r}{\mathbf{r}}

\newcommand{\diag}{\textup{diag}}
\newcommand{\G}{\mathbf{G}}

\long\def\symbolfootnote[#1]#2{\begingroup\def\thefootnote{\fnsymbol{footnote}}
\footnote[#1]{#2}\endgroup}

\begin{document}
%
\title{OpenRANet: Neuralized Spectrum Access by Joint Subcarrier and Power Allocation with Optimization-based Deep Learning
}
%
%
%

 \author{Siya Chen, Chee Wei Tan, Xiangping Zhai, and H. Vincent Poor\
  \thanks{S. Chen is with the Department of Computer Science, City University of Hong Kong, Hong Kong (e-mail: siyachen4-c@my.cityu.edu.hk).}
  \thanks{C. W. Tan is with Nanyang Technological University, Singapore, Nanyang Ave., Singapore (e-mail: cheewei.tan@ntu.edu.sg).}
 \thanks{X. B. Zhai is with the College of Computer Science and Technology, Nanjing University of Aeronautics and Astronautics, Nanjing 211106, China, and also with the Collaborative Innovation Center of Novel Software Technology and Industrialization, Nanjing 210023, China (e-mail: blueicezhaixp@nuaa.edu.cn).}
 \thanks{H. V. Poor is with the Department of Electrical and Computer Engineering, Princeton University, Princeton, NJ 08544 USA (e-mail: poor@princeton.edu).}
 \thanks{The material in this paper was presented in part at the Global Communications Conference (GLOBECOM), Singapore, 2017.}
\thanks{This work is supported in part by the Jiangsu Province Frontier Leading Technology Basic Research Project under Grant BK20222013, the U.S National Science Foundation under Grant ECCS-2335876, and the Singapore Ministry of Education Academic Fund (RG91/22).}
 }

\maketitle
\thispagestyle{empty}

\begin{abstract}
\textcolor{black}{
The next-generation radio access network (RAN), known as Open RAN, is poised to feature an AI-native interface for wireless cellular networks, including emerging satellite-terrestrial systems, making deep learning integral to its operation. In this paper, we address the nonconvex optimization challenge of joint subcarrier and power allocation in Open RAN, with the objective of minimizing the total power consumption while ensuring users meet their transmission data rate requirements. We propose OpenRANet, an optimization-based deep-learning model that integrates machine-learning techniques with iterative optimization algorithms. We start by transforming the original nonconvex problem into convex subproblems through decoupling, variable transformation, and relaxation techniques. These subproblems are then efficiently solved using iterative methods within the standard interference function framework, enabling the derivation of primal-dual solutions. These solutions integrate seamlessly as a convex optimization layer within OpenRANet, enhancing constraint adherence, solution accuracy, and computational efficiency by combining machine learning with convex analysis, as shown in numerical experiments. OpenRANet also serves as a foundation for designing resource-constrained AI-native wireless optimization strategies for broader scenarios like multi-cell systems, satellite-terrestrial networks, and future Open RAN deployments with complex power consumption requirements.
}
\end{abstract}


%
\IEEEpeerreviewmaketitle

\section{Introduction}\label{introduction}

\textcolor{black}{The Open Radio Access Network (Open RAN) architecture, characterized by its open interfaces, decentralized network elements, virtualized hardware and software, and intelligent control, represents a transformative approach to enhancing network deployment, fostering innovation and reducing costs for next-generation wireless networks \cite{oran2018,polese2023understanding,abdalla2022toward}. Open RAN can also integrate satellite-terrestrial networks (e.g., SpaceX Starlink), significantly extending its reach and improving connectivity in remote and underserved areas \cite{nvidia1,spacex3,spacex4}, thereby enhancing overall network performance and user experience. Furthermore, Open RAN can leverage deep learning technologies to embed intelligence into every unit of the architecture \cite{polese2023understanding,abdalla2022toward}. By employing deep learning techniques, Open RAN can achieve intelligent decision-making at various levels, leading to improved spectrum efficiency, reduced power consumption, dynamic resource allocation, optimized network coverage, and decreased operational costs. This integration of Open RAN, satellite-terrestrial networks, and deep learning enhances adaptability, efficiency, and scalability, making these systems well-suited to the demands of 6G and beyond \cite{polese2023understanding,abdalla2022toward}. However, it is essential to assess the technology's impact on energy consumption to ensure that the benefits of Open RAN are not negated by increased energy usage.
}

\textcolor{black}{
 The impact of Open RAN on network energy consumption is critical, influencing operational costs, environmental sustainability, and overall network performance \cite{abdalla2022toward,polese2023understanding}. With its decentralized and virtualized network elements, Open RAN offers opportunities to optimize energy usage compared to traditional RAN systems. Implementing intelligent power management techniques, which dynamically adjust the power levels of network components based on real-time traffic demands, is one approach to reducing energy consumption. In this paper, we focus on optimizing power allocation in Open RAN systems to minimize total power consumption while ensuring that each user meets the required transmission data rates, thereby guaranteeing that the total transmission data rate from all subcarriers for each user meets or exceeds a predefined threshold.}

Extensive research has been conducted on optimizing power usage in wireless networks \cite{tan2008energy,fu2020zero,pennanen2015decentralized,zheng2017max,luo2019deep,tan2011spectrum,tan2015optimal,zhai2013energy}, covering various scenarios such as MISO \cite{fu2020zero}, MIMO \cite{pennanen2015decentralized}, multi-carrier \cite{luo2019deep}, and multi-cell \cite{pennanen2015decentralized} systems. These works utilized optimization-based algorithms to minimize power consumption while accommodating different data rate constraints. For instance, Pennanen et al. \cite{pennanen2015decentralized} employed a successive convex approximation method (SCA) to solve the power minimization problem with per-user rate constraints in a multi-cell multi-user MIMO system. 
Fu et al. \cite{fu2020zero} proposed a two-stage algorithm named power consumption-based user clustering to address the power consumption minimization problem in a generic multi-cell system. However, these optimization-based algorithms are often prone to getting trapped in local optima and may have high computational costs, limiting their scalability as the network size increases. Addressing nonconvexity in a distributed manner thus remains a challenge.

Open RAN's AI-native layer enables the integration of advanced technologies like deep learning and big data analytics, opening new avenues for addressing nonconvexity in power consumption problems. This approach reduces energy usage, lowers operational costs, and enhances network performance.  Recent works have also utilized machine learning techniques to learn optimal power allocation strategies for reducing power consumption based on observed data for the wireless systems \cite{luo2019deep, camana2022deep, park2023deep}. Park et al. \cite{park2023deep} proposed a power control algorithm for sub-band assignment and unsupervised deep neural networks in IoT networks. Similarly, the work in \cite{luo2019deep} used deep learning for minimizing total transmit power in simultaneous wireless information and power transfer systems. Camana et al. \cite{camana2022deep} used a deep neural network-based convex relaxation technique to reduce transmission power in multi-user MISO-SWIPT systems. Machine learning techniques have also advanced other wireless network optimization problems, including sum rate maximization \cite{eisen2020optimal, li2021toward,chen2023neural}, energy efficiency maximization \cite{giannopoulos2021deep, chuang2023deep}, user fairness maximization \cite{jang2022alpha}, and secrecy rate maximization \cite{sharma2023secrecy}. However, as discussed in \cite{o2017introduction,song2022benchmarking}, machine learning techniques are unlikely to completely replace resource allocation algorithms in communication systems, which are often designed from an optimization theory perspective. Also, machine learning methods often require extensive parameters to model complex systems, leading to substantial training data and costs. By integrating modern machine-learning approaches with traditional optimization-based techniques, there is significant potential to enhance both the effectiveness and efficiency of addressing the complexity challenges in wireless network optimization \cite{o2017introduction,song2022benchmarking,chen2024neural}.

This paper tackles the non-convex problem of joint subcarrier and power control to minimize total power while meeting rate requirements. The complexity arises from non-convexity, coupling constraints, and implicit resource uncertainties. We propose OpenRANet, an optimization-based deep learning model for effective subcarrier and power allocation in open RAN systems, specifically by incorporating a convex optimization layer into deep learning models \cite{agrawal2019differentiable, agrawal2020differentiating}. Our approach decomposes the non-convex problem into solvable convex subproblems, which are then integrated as convex optimization layers within the deep learning model.  We start by extracting convex subproblems from the non-convex original problem using decoupling, change-of-variable strategies, and convex relaxation. We then apply the standard interference function framework to derive the primal and dual solutions of these subproblems. These solutions are used as a convex optimization layer in OpenRANet, which integrates deep learning techniques like convolutional filters and neural networks.  During training, the model alternates between solving these convex subproblems and updating other parameters based on the solutions. This integration of convex optimization layers allows deep learning models to leverage iterative optimization techniques, improving efficiency, accuracy, and convergence in tackling non-convex optimization problems \cite{boydcvxmodel1, boydcvxmodel2, amos2023tutorial}.


The novelty of OpenRANet lies in its integration of conventional optimization methods with deep-learning training, allowing for the optimization of the optimizer itself by adjusting parameters and hyperparameters, thus reducing manual tuning effort. Broadly, OpenRANet can be categorized as a learned optimizer \cite{boydcvxmodel1,boydcvxmodel2,amos2023tutorial}, which transforms traditional optimization models into differentiable machine learning models and facilitates solving nonconvex resource allocation problems in a decentralized manner. Overall, the contributions of the paper are as follows:
\begin{enumerate}

\item \textcolor{black}{First, we consider the nonconvex problem of minimizing total transmitted power consumption subject to transmission rate requirements for multi-subcarrier open RAN systems. We extract the convex subproblems from the original nonconvex problem using decoupling techniques, change-of-variable strategies, and convex relaxation. Then, by leveraging the log-convexity property of the standard interference function and insights from Lagrangian duality, we propose a low-complexity iterative primal-dual algorithm to achieve at least a locally optimal solution.}
\textcolor{black}{
\item Secondly, we propose an optimization-based deep learning model, named OpenRANet, which incorporates projection and convex optimization layers to address the issue of the primal-dual algorithm converging to local optima. The mechanism of OpenRANet integrates the analysis of the convex subproblem and the constraint information of the nonconvex original problem into deep learning techniques. These integrations enable the OpenRANet algorithm to excel in adhering to problem constraints, significantly enhancing solution accuracy compared to pure machine learning approaches, and greatly improving computational efficiency over optimization-based methods.
}
\textcolor{black}{
\item Finally, we evaluate the performance of the proposed OpenRANet in approximating global optima for the power minimization problem, comparing it with other state-of-the-art optimization-based methods and machine learning models to demonstrate its effectiveness and efficiency. The OpenRANet framework can also serve as a foundation for AI-native wireless optimization strategies in broader scenarios, including multi-cell systems and future open RAN systems with additional power consumption requirements.}

\end{enumerate}

The rest of the paper is organized as follows:
First, we introduce the formulation of the nonconvex total power minimization problem for the open RAN systems with multiple subcarriers and transmission rate requirements in Section~\ref{sec:model}. 
Next, we examine the standard interference function framework and the Lagrangian duality to design low-complexity iterative algorithms in Section~\ref{sec:fixpoint}. 
Then, in Section~\ref{sec:CVXL}, we present the proposed OpenRANet, an optimization-based learning model, to approximate the optimal solution to the non-convex problem. 
In Section~\ref{sec:experiments}, we present numerical simulations. 
Finally, Section~\ref{sec:conclusion} concludes the paper.

\section{System Model}\label{sec:model}

\textcolor{black}{
We consider a downlink multi-carrier open RAN system, comprising a single base station and $L$ active users who share $M$ subcarriers. In this scenario, the base station operates as the transmitter, dispatching signals to the receivers, represented by the $L$ active users across $L$ unique communication links.  We denote the index set of users by $l\in \{1,..., L\}$, and the set of subcarriers by $m \in \{1,...,M\}$. We assume the total bandwidth is $B$ and each subcarrier has a bandwidth of $B/M$.
In addition, we assume that the number of active users can exceed the number of subcarriers and that multiple receivers may share a single subcarrier, potentially leading to interference between receivers. Each receiver is capable of utilizing the necessary number of subcarriers to satisfy the bit rate requirements of its particular application. 
Let $p_l^m$ denote the transmit power from the $\mathrm{BS}$ to  $l$-th user on the $m$-th subcarrier. \textcolor{black}{User $l$ is said to be active on subcarrier $m$ if $p_l^m > 0$, and inactive otherwise}. Let $G_{lj}^m$ stands for the channel gain between the $j$-th and the $l$-th user through the $m$-th subcarrier, and $\sigma^m_l$ is the additive white Gaussian noise. We assume that the channel gains are perfectly known. The system model is illustrated in Fig. \ref{fig:wf}.
Treating interference as noise, the ${\mathsf{SINR}}$ of the $l$-th receiver on the $m$-th subcarrier can be expressed in terms of $\mathbf{p}^m=\left(p_1^m,p_2^m,\ldots,p_L^m\right)^{\top}$ as \cite{tan2015wireless}
\begin{align*}
{\mathsf{SINR}}_l^m(\mathbf{p}^m)=\frac{G_{ll}^mp_l^m}{\sum_{j\neq l}G_{lj}^mp_j^m+\sigma^m_l}.
\end{align*}
}

\begin{figure}
\centerline{\includegraphics[scale=0.36]{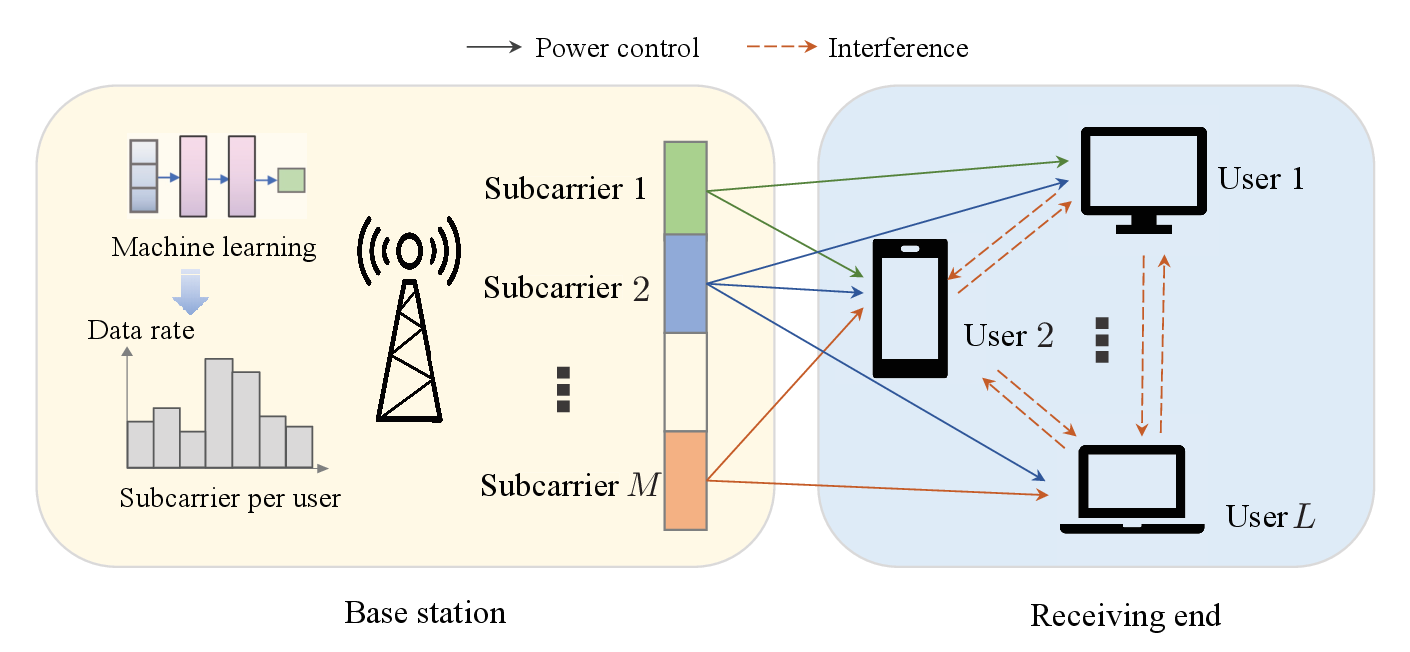}}
\captionsetup{font={footnotesize}, justification=raggedright}
\caption{ The architecture of the system model for a downlink multi-carrier open RAN system with $L$ active users and $M$ subcarriers.}
\label{fig:wf}
\end{figure}

\begin{table*}[t]
\centering
\begin{tabular}{| p{2.8cm} || c | c | c | }
\hline
 & CDMA Approximation&Shannon Capacity&Bit Error Rate\\
\hline \hline
Rate Functions:$\quad\quad\;$ $f_l^m(\scalemath{0.9}{\mathsf{SINR}}_l^m(\mathbf{p}^m))$& $\scalemath{0.9}{\mathsf{SINR}}_l^m(\mathbf{p}^m)$ & $\log (1+\scalemath{0.9}{\mathsf{SINR}}_l^m(\mathbf{p}^m))$ & $R(1-2Q(\sqrt{\scalemath{0.9}{\mathsf{SINR}}_l^m(\mathbf{p}^m)}))$\\
\hline
Standard Interference Functions: $I_l^m(\mathbf{p}^m)$ & $\displaystyle\frac{r_l^m}{\scalemath{0.9}{\mathsf{SINR}}_l^m(\mathbf{p}^m)}p_l^m$ & $\displaystyle \frac{r_l^m}{\log (1+\scalemath{0.9}{\mathsf{SINR}}_l^m(\mathbf{p}^m))}p_l^m$ & $\displaystyle \frac{r_l^m}{R(1-2Q(\sqrt{\scalemath{0.9}{\mathsf{SINR}}_l^m(\mathbf{p}^m)}))}p_l^m$\\
\hline
Log-convexity: $\log I_l^m(e^{\tilde{\mathbf{p}}^m})$& $\log \displaystyle\frac{r_l^m}{\scalemath{0.9}{\mathsf{SINR}}_l^m(e^{\tilde{\mathbf{p}}^m})}e^{\tilde{p}_l^m}$ & $ \log \displaystyle \frac{r_l^m}{\log (1+\scalemath{0.9}{\mathsf{SINR}}_l^m(e^{\tilde{\mathbf{p}}^m}))}e^{\tilde{p}_l^m}$ & $\log \displaystyle \frac{r_l^m}{R(1-2Q(\sqrt{\scalemath{0.9}{\mathsf{SINR}}_l^m(e^{\tilde{\mathbf{p}}^m})}))}e^{\tilde{p}_l^m}$\\
\hline 
\end{tabular}
\\[1ex]
\captionsetup{font={footnotesize}, justification=raggedright}
\caption{Table of Heterogeneous Rate Functions for Interference Function Modeling}
\label{table:ratefunction}
\end{table*}



The problem of minimizing the total power consumption subject to given transmission rate requirements with multiple subcarriers is formulated as follows:
\begin{align}\label{opt:power_mnos}
    \text{minimize}&\ \displaystyle \sum_{l=1}^L\sum_{m=1}^M p_l^m \nonumber\\
    \text{subject to}&\ \displaystyle \sum_{m=1}^M f_l^m(\scalemath{0.9}{\mathsf{SINR}}_l^m(\mathbf{p}^m))\!\geq\!\bar{r}_l\quad\forall \;l,\\
    \text{variables:}&\ \mathbf{p}^m=(p_1^m,\cdots,p_L^m)^\top\!\geq\!\mathbf{0}\quad\forall \;m,\nonumber
\end{align}
where $f_l^m(\scalemath{0.9}{\mathsf{SINR}}_l^m(\mathbf{p}^m))$ is the achievable transmission rate function of the $l$-th receiver through the $m$-th subcarrier, as defined in TABLE \ref{table:ratefunction}, and $\bar{r}_l>0$ is the minimum transmission rate requirement of the $l$-th user.
In general, it may not be possible to have the transmission rates of all users be higher than $\bar{\mathbf{r}}=(\bar r_1,\bar r_2,\ldots,\bar r_L)^\top$ in~\eqref{opt:power_mnos}.
Here, we assume that~\eqref{opt:power_mnos} is feasible for the given $\bar{\mathbf{r}}$.
We also assume that the function $f_l^m(\scalemath{0.9}{\mathsf{SINR}}_l^m(\mathbf{p}^m))$ that models any kind of transmission rate in a general wireless network is doubly differentiable.

The problem in~\eqref{opt:power_mnos} is challenging to solve due to the nonconvexity in the transmission rate constraints since $f_l^m(\scalemath{0.9}{\mathsf{SINR}}_l^m(\mathbf{p}^m))$ for all $l$ and $m$ are in general nonlinear and nonconvex functions of the powers.
In addition, a practical requirement is that the power solution of each user is determined with minimal coordination between the multiple subcarriers.
We proposed two steps to solve \eqref{opt:power_mnos}. The first step is to exploit a fixed-point characterization of the primal constraints associated with the transmission rate requirements in~\eqref{opt:power_mnos}.
This leads to the design of a low-complexity iterative primal-dual algorithm.
The second step is to design and train an optimization-based deep learning model in the cloud core by leveraging jointly convolutional neural networks and convex optimization models, which the iterative primal-dual algorithm can solve in the first step. 



\section{Reweighted Primal-dual Algorithm for Total Power Minimization}\label{sec:fixpoint}
In this section, we reformulate and decouple the total power consumption minimization problem in \eqref{opt:power_mnos}. We analyze the constraints of the reformulated problem, which can be interpreted as log-convex standard interference functions. Then we propose a primal-dual algorithm to obtain the local optimal solution to the problem 
 in \eqref{opt:power_mnos} by leveraging the properties of the log-convex standard interference functions.

\subsection{Reformulation of the Total Power Minimization}

We first introduce the auxiliary rate variables $\mathbf{r}^m=(r_1^m,$ $r_2^m,\ldots,r_L^m)^{\top}$ for all $m$, and rewrite \eqref{opt:power_mnos} as
\begin{align}\label{opt:power_mnos_global}
    \text{minimize}&\ \displaystyle \sum_{l=1}^L\sum_{m=1}^M p_l^m\nonumber \\
    \text{subject\;to}&\ \displaystyle f_l^m(\scalemath{0.9}{\mathsf{SINR}}_l^m(\mathbf{p}^m))\!\geq\!r_l^m \quad\forall\; l,\;\forall\; m,\nonumber \\
    &\ \displaystyle \sum_{m=1}^M r_l^m \geq \bar{r}_l\quad\forall \;l, \\
    \text{variables:}&\ \mathbf{p}^m\geq\! \mathbf{0},\;
     \mathbf{r}^m\!\geq\! \mathbf{0}\quad\forall \;m.\nonumber
\end{align}

The problem in \eqref{opt:power_mnos_global} allows us to decouple the transmission rate constraints across the multiple subcarriers. Now let us suppose the problem in \eqref{opt:power_mnos_global} or \eqref{opt:power_mnos} is feasible. 
Denote the optimal transmission rate for each subcarrier by ${\mathbf{r}^m}^\star$ and the optimal power allocation by ${\p^{m*}}$. Then we have the following theorem.

\begin{theorem}\label{theorem2}
If the problem in \eqref{opt:power_mnos_global} is feasible, then the transmission rate constraints in~\eqref{opt:power_mnos_global} are tight when \eqref{opt:power_mnos_global} reaches to its optima, which means
\begin{equation*}
\displaystyle \sum_{m=1}^M {r_l^m}^\star= \sum_{m=1}^M f_l^m(\scalemath{0.9}{\mathsf{SINR}}_l^m({\mathbf{p}^m}^\star)) = \bar{r}_l\quad\forall \;l.\\
\end{equation*}
\end{theorem}

Now we introduce the standard interference function in \cite{yates1995framework}, which motivates us to design an iterative primal-dual algorithm and provides a convergence guarantee for our algorithm.
\begin{definition}\cite{yates1995framework}\label{def:sif}
 $\mathbf{I}(\mathbf{p})$ is a standard interference function if, for all $\mathbf{p}\geq 0$, the following properties are satisfied:
\begin{enumerate}
    \item (Positivity) $\mathbf{I}(\mathbf{p})\geq \bm{0}$.
    \item (Monotonicity) If $\mathbf{p}_1\geq \mathbf{p}_2$, then $\mathbf{I}(\mathbf{p}_1)\geq \mathbf{I}(\mathbf{p}_2)$.
    \item (Scalability) For any $\alpha>1, \alpha\mathbf{I}(\mathbf{p})>\mathbf{I}(\alpha\mathbf{p})$.
\end{enumerate}
\end{definition}

The main convergence result for standard interference functions can be summarized as follows:

\begin{lemma}\cite{yates1995framework}\label{lem:lem1}
Given a standard interference function $\mathbf{I}(\mathbf{p})$, if there exists a $\p'\geq \bm{0}$ such that $\mathbf{I}(\mathbf{p}')\!\leq\!\p'$, then $\mathbf{I}(\mathbf{p})$ has a fixed point, i.e., there exists a $\p^\star$ such that $\mathbf{p}^\star\!=\!\mathbf{I}(\mathbf{p}^\star)$. Moreover, the iteration $\p(k\!+\!1)\!=\!\mathbf{I}(\mathbf{p}(k))$ converges to the fixed-point from any initial point $\p(0)$.
\end{lemma}

Then for the transmission rate function  $f_l^m(\scalemath{0.9}{\mathsf{SINR}}_l^m(\mathbf{p}^m))$, we can construct an interference function by the following lemma.
\begin{lemma}\label{p_standard}
For all $ l$ and all $m$, if the transmission rate function $f_l^m(\scalemath{0.9}{\mathsf{SINR}}_l^m(\mathbf{p}^m))$ is positive and satisfies 
\begin{equation}\label{eq:scal}
\displaystyle \frac{\partial f_l^m(\scalemath{0.9}{\mathsf{SINR}}_l^m(\mathbf{p}^m))}{\partial \scalemath{0.9}{\mathsf{SINR}}_l^m(\mathbf{p}^m)}>0, \nonumber
\end{equation}
and
\begin{equation}\label{eq:der1}
 \displaystyle \frac{\partial f_l^m(\scalemath{0.9}{\mathsf{SINR}}_l^m(\mathbf{p}^m))}{\partial\scalemath{0.9}{\mathsf{SINR}}_l^m(\mathbf{p}^m)} \leq \frac{f_l^m(\scalemath{0.9}{\mathsf{SINR}}_l^m(\mathbf{p}^m))}{\scalemath{0.9}{\mathsf{SINR}}_l^m(\mathbf{p}^m))}, \nonumber
\end{equation}
then by letting 
\begin{equation}\label{eq:sif}
I_l^m(\mathbf{p}^m)=\displaystyle \frac{{r_l^m}}{f_l^m(\scalemath{0.9}{\mathsf{SINR}}_l^m(\mathbf{p}^m)}p_l^m,
\end{equation}
$\mathbf{I}^m(\mathbf{p}^m)=(I_1^m(\mathbf{p}^m),I_2^m(\mathbf{p}^m),\cdots,I_L^m(\mathbf{p}^m))^\top$ is a standard interference function with respect to $\p^m$ for all $m$. 
\end{lemma}

\begin{corollary}\label{corol1}
For the transmission rate functions given in Table \ref{table:ratefunction}, the interference function \eqref{eq:sif} is standard and $I_l^m(\mathbf{p}^m)=\mathbf{p}^m$ has a fixed-point for all $l,m$.  
\end{corollary}

More details of the proof of Corollary \ref{corol1}  are available in \href{https://arxiv.org/abs/2409.12964} {the ArXiv online version}.
\textcolor{black}{
\begin{remark}
Lemma \ref{p_standard} establishes sufficient conditions under which a standard interference function can be constructed from the transmission rate function. Consequently, for those transmission rate functions that adhere to these conditions, we can leverage the properties of the standard interference function to analyze and deal with the constraints related to the transmission rate functions of the problem in \eqref{opt:power_mnos_global}. Beyond the transmission rate function defined in TABLE \ref{table:ratefunction}, our framework is also applicable to the problem \eqref{opt:power_mnos_global} with any transmission rate function constraint that complies with the conditions in Lemma~\ref{p_standard}.
\end{remark}
}

\begin{definition}\label{def:logconvex}
$\I(\mathbf{p})$ is a \textit{log-convex standard interference function} if it fulfills the condition of standard interference function~\cite{yates1995framework} and, in addition, in the logarithmic domain, i.e., $\tilde{\mathbf{p}}=\log \mathbf{p}$, $\I(e^{\tilde{\mathbf{p}}})$ is log-convex in terms of $\tilde{\p}$.
\end{definition}

\begin{lemma}\label{thm3}
If the transmission rate function $f_l^m(\scalemath{0.9}{\mathsf{SINR}}_l^m(\mathbf{p}^m))$ satisfies that
 \begin{align}\label{eq:thm3_1}
&\left(\frac{\partial^2 f_l^m(\scalemath{0.9}{\mathsf{SINR}}_l^m(\mathbf{p}^m))}{\partial \scalemath{0.9}{\mathsf{SINR}}_l^m(\mathbf{p}^m)^2}\scalemath{0.9}{\mathsf{SINR}}_l^m(\mathbf{p}^m)+\frac{\partial f_l^m(\scalemath{0.9}{\mathsf{SINR}}_l^m(\mathbf{p}^m))}{\partial \scalemath{0.9}{\mathsf{SINR}}_l^m(\mathbf{p}^m)}\right)\nonumber\\
& \cdot f_l^m(\scalemath{0.9}{\mathsf{SINR}}_l^m(\mathbf{p}^m)) \leq \left(\frac{\partial f_l^m(\scalemath{0.9}{\mathsf{SINR}}_l^m(\mathbf{p}^m))}{\partial \scalemath{0.9}{\mathsf{SINR}}_l^m(\mathbf{p}^m)}\right)^2\scalemath{0.9}{\mathsf{SINR}}_l^m(\mathbf{p}^m),
\end{align}
then it is log-concave in terms of $\tilde{\mathbf{p}}^m=\log \mathbf{p}^m$ for all $l$, $m$.
\end{lemma}

\begin{corollary}\label{corol2}
The transmission rate function $f_l^m(\scalemath{0.9}{\mathsf{SINR}}_l^m(\mathbf{p}^m))$ defined in TABLE \ref{table:ratefunction} are all log-concave in terms of $\tilde{\mathbf{p}}^m=\log \mathbf{p}^m$ for all $l$ and $m$.
\end{corollary}

More details of the proof of Corollary \ref{corol2}  are available in \href{https://arxiv.org/abs/2409.12964} {the ArXiv online version}.
\textcolor{black}{
\begin{remark}
    Lemma \ref{thm3} further provides a sufficient condition for constructing a concave transmission rate function through the logarithmic change-of-variable technique. In fact, with this logarithmic change-of-variable technique, the interference function \eqref{eq:sif} in Lemma \ref{p_standard} is a \textit{log-convex standard interference function}, as we have 
\begin{equation*}\label{extension}
\displaystyle \log I_l^m(e^{\tilde{\mathbf{p}}^m})=\log r_l^m + \tilde{p}_l^m - \log (f_l^m(\scalemath{0.9}{\mathsf{SINR}}_l^m(e^{\tilde{\mathbf{p}}^m})))
\end{equation*} for all $l$ and $m$, which is convex based on Lemma~\ref{thm3}.
The log-concavity of the transmission rate function is important because the log-convex standard interference function derived from it can be used to build iterative algorithms for solving subproblems of  \eqref{opt:power_mnos_global}. It also facilitates the use of optimization theory to derive the dual of the problem. This is crucial for applying the convex optimization layer to construct deep learning models, which we will elaborate on in the next section.
\end{remark}
}


\subsection{Reweighted Primal-dual Algorithm for Total Power Minimization}

Next, we give another reformulation to~\eqref{opt:power_mnos_global}. By letting $\tilde{p}_l^m=\log p_l^m$ and $\tilde{r}_l^m=\log r_l^m$ for all $l$ and $m$, and taking the logarithm on both sides of the individual rate constraints in~\eqref{opt:power_mnos_global},
we then rewrite~\eqref{opt:power_mnos_global} as the following equivalent optimization problem:
\begin{align}\label{opt:power_mno_alt}
    \text{minimize}&\ \displaystyle \sum_{l=1}^L\sum_{m=1}^M e^{\tilde{p}_l^m} \nonumber\\
    \text{subject to}&\ \displaystyle \log f_l^m(\scalemath{0.9}{\mathsf{SINR}}_l^m(e^{\tilde{\mathbf{p}}^m})) \geq \tilde{r}_l^m\quad\forall \;l,\ \forall \; m,\\
    &\ \displaystyle \sum_{m=1}^M e^{\tilde{r}_l^m} \geq \bar{r}_l\quad\forall \;l, \nonumber\\
    \text{variables:}&\ \tilde{\p}^m, \;
     \tilde{\r}^m\quad\forall \;m. \nonumber
\end{align}

Suppose the transmission rate function $f_l^m(\scalemath{0.9}{\mathsf{SINR}}_l^m(\mathbf{p}^m))$ satisfies the condition in Lemma~\ref{thm3}, then the nonconvexity associated with~\eqref{opt:power_mno_alt} lies in the $L$ transmission rate requirement constraints:
\begin{equation}\label{nonconvexity}
\displaystyle \sum_{m=1}^M e^{\tilde{r}_l^m}\! \geq\! \bar{r}_l\quad\forall \;l.
\end{equation}
\textcolor{black}{
 This is because $e^{\tilde{r}_l^m}$ is convex, so the feasible set defined by these constraints is not convex. Consequently, traditional optimization techniques, such as the interior point method, are not applicable for solving the problem in \eqref{opt:power_mno_alt}.
Except for~\eqref{nonconvexity}, all the other constraint sets and the objective function in~\eqref{opt:power_mno_alt} are convex.}

We now consider formulating a relaxed subproblem of ~\eqref{opt:power_mno_alt}. Specifically, if we ignore  \eqref{nonconvexity}, then \eqref{opt:power_mno_alt} can be relaxed as 
\begin{align}\label{opt:power_relaxed}
    \text{minimize}&\ \displaystyle \sum_{l=1}^L\sum_{m=1}^M e^{\tilde{p}_l^m} \nonumber\\
    \text{subject to}&\ \displaystyle \log f_l^m(\scalemath{0.9}{\mathsf{SINR}}_l^m(e^{\tilde{\mathbf{p}}^m})) \geq \tilde{r}_l^m\quad\forall \;l,\ \forall \ m,\\
    \text{variables:}&\ \tilde{\p}^m, \; \tilde{\r}^m\quad\forall \;m. \nonumber
\end{align}
The Lagrangian duality of this relaxed problem in \eqref{opt:power_relaxed} (in the logarithmic domain) provides a way to assign a user to a less-congested subcarrier for rate allocation using interpreting the Lagrangian dual solution as the price to pay to meet the rate constraints in the presence of the other users. The Lagrange duality also serves as a way to decouple and decentralize the resource allocation over the spectrum.  

Suppose that we are given a set of individual transmission rate requirements $\hat{r}_l^m$ in the feasible region, i.e. $\sum_{m=1}^M e^{\hat{r}_l^m}\! \geq\! \bar{r}_l$ for all $l$. Then with the given $\hat{r}_l^m$ in \eqref{opt:power_relaxed}, we have the following convex subproblem

\begin{align}\label{opt:power_mno_cvx}
    \text{minimize}&\ \displaystyle \sum_{l=1}^L\sum_{m=1}^M e^{\tilde{p}_l^m}\nonumber\\
    \text{subject to}&\ \displaystyle \log f_l^m(\scalemath{0.9}{\mathsf{SINR}}_l^m(e^{\tilde{\mathbf{p}}^m})) \geq \hat{r}_l^m \quad \forall\;l,\forall\;m, \nonumber\\
    \text{variables:}&\ \tilde{\p}^m=(\tilde{p}_1^m,\cdots,\tilde{p}_L^m)^\top\quad\forall \;m.
\end{align}

Note that~\eqref{opt:power_mno_cvx} is strictly convex, and we assume that there exist strictly feasible powers that satisfy Slater's condition~\cite{Convex_Optimization}.
In particular, strong duality holds in~\eqref{opt:power_mno_cvx}. According to the Karush-Kuhn-Tucker (KKT) optimality conditions of~\eqref{opt:power_mno_cvx}, we have the following theorem to solve \eqref{opt:power_mno_cvx} efficiently.

\begin{theorem}\label{th:dual}
By introducing an auxiliary variable
$x_l^m$ as
\begin{align*}
    x_l^m = \frac{\lambda_l^m \scalemath{0.9}{\mathsf{SINR}}_l^m({\mathbf{p}^m})}{p^m_lf_l^m(\scalemath{0.9}{\mathsf{SINR}}_l^m({\mathbf{p}^m}))}\frac{\partial f_l^m(\scalemath{0.9}{\mathsf{SINR}}_l^m({\mathbf{p}^m}))}{\partial \scalemath{0.9}{\mathsf{SINR}}_l^m({\mathbf{p}^m})},
\end{align*}
the Lagrangian multiplier ${\boldsymbol{\lambda}^m}^\star$ corresponding to the rate constraints in~\eqref{opt:power_mno_cvx}, the optimal transmit power ${\mathbf{p}^m}^\star$ and rate ${\mathbf{r}^m}^\star$ in~\eqref{opt:power_mnos_global} for all $l$ and $m$ satisfy
\begin{align}\label{eq:thm4_1}
 f_l^m(\scalemath{0.9}{\mathsf{SINR}}_l^m({\mathbf{p}^m}^\star))={r_l^m}^\star,
\end{align}
\begin{align*}
 {x_l^m}^\star =1+\displaystyle \sum_{j\neq l}\frac{G_{lj}^m}{G_{ll}^m}\scalemath{0.9}{\mathsf{SINR}}_j^{m}({\mathbf{p}^m}^\star){x_j^m}^\star,
\end{align*}
and
\begin{align*}
{\lambda_l^m}^\star=\frac{f_l^m(\scalemath{0.9}{\mathsf{SINR}}_l^m({\mathbf{p}^m}^\star))}{\scalemath{0.9}{\mathsf{SINR}}_l^m({\mathbf{p}^m}^\star)\frac{\displaystyle\partial f_l^m(\scalemath{0.9}{\mathsf{SINR}}_l^m({\mathbf{p}^m}^\star))}{\displaystyle\partial \scalemath{0.9}{\mathsf{SINR}}_l^m({\mathbf{p}^m}^\star)}}{p_l^m}^\star {x_l^m}^\star.
\end{align*}
Furthermore, $\mathbf{p}^{m*}$ and $\mathbf{x}^{m*}$ can be obtained by the following iterations with any initial $\mathbf{p}^{m}(0)$ and $\mathbf{x}^{m}(0)$:
\begin{align*} 
    {p_l^m}(k+1)\leftarrow \frac{e^{\hat{r}_l^m}}{f_l^m(\scalemath{0.9}{\mathsf{SINR}}_l^m(\mathbf{p}^m)}p_l^m
    (k),
\end{align*}
\begin{align*}
    x_l^m(k+1)\leftarrow1+\sum_{j\neq l}\frac{G_{lj}^m}{G_{ll}^m}\scalemath{0.9}{\mathsf{SINR}}(t)_j^{m}x_j^m(k).
\end{align*}
\end{theorem}

\textcolor{black}{
\begin{remark}
    The primal solution of~\eqref{opt:power_mno_cvx} serves as a feasible solution for~\eqref{opt:power_mno_alt}.  When the individual transmission rate requirements are at their optimal, that is, $e^{\hat{r}_l^m}=r_l^{m^\star}$ for all $l$ and $m$, the solution of~\eqref{opt:power_mno_cvx} becomes the optimal solution for~\eqref{opt:power_mno_alt}. The iterations for ${p_l^m}(k+1)$ and $ x_l^m(k+1)$ are assured to converge due to the properties of the standard interference function. The Lagrangian dual ${\boldsymbol{\lambda}^m}^\star$ can subsequently be acquired using the fixed points of the iterations in Therem \ref{th:dual}. 
As previously discussed, the Lagrangian dual is important for the application of the convex optimization layer in developing deep learning models. 
\end{remark}
} 

The advantage of considering~\eqref{opt:power_mno_cvx} is to determine the parameter at each subcarrier by leveraging the Lagrangian duality in~\eqref{opt:power_mno_cvx}, i.e., to compute the Lagrangian dual solution $\boldsymbol{\lambda}^m$. Intuitively, the total power consumption can be minimized by minimizing the multiuser interference at each subcarrier whenever possible, i.e., when users are admitted, a user should choose a less-congested subcarrier by using the congested information captured by $\lambda_l^m$ for the $l$-th user on the $m$-th subcarrier. The power consumption can be minimized by mitigating the congestion among users at the subcarrier based on the Lagrangian dual parameter $\lambda_l^m$ using Algorithm \ref{alg:decentralizedAlg}.

\begin{algorithm}
\caption{Reweighted Primal-dual Algorithm}
\label{alg:decentralizedAlg}
\SetAlgoLined
\textbf{Initialize}: $\mathbf{\omega}_l^m(0)\geq 0$ with $\sum_{m=1}^M \mathbf{\omega}_l^m(0)=1$, $\textup{tol}>0$ and  $\epsilon>\textup{tol}$.\\
\begin{enumerate}
\item Obtain $l$-th user's rate with $m$-th subcarrier:
\begin{align*}
    r_l^m(t)\leftarrow\omega_l^m(t)\bar{r}_l.
\end{align*}

\item Obtain $\mathbf{p}^{m*}(t)$ and $\mathbf{x}^{m*}(t)$ by the following iterations with any initial $\mathbf{p}^{m}(0)$ and $\mathbf{x}^{m}(0)$ until convergence:
\begin{align}\label{eq:iter_p}  
    {p_l^m}(k+1)\leftarrow \frac{e^{\hat{r}_l^m}}{f_l^m(\scalemath{0.9}{\mathsf{SINR}}_l^m(\mathbf{p}^m)}p_l^m
    (k),
\end{align}
\begin{align}\label{eq:virtual_x}
    x_l^m(k+1)\leftarrow1+\sum_{j\neq l}\frac{G_{lj}^m}{G_{ll}^m}\scalemath{0.9}{\mathsf{SINR}}(t)_j^{m}x_j^m(k).
\end{align}
\item Obtain the Lagrangian multiplier $\bm{\lambda}^{m*}(t)$ by:
\begin{align}\label{eq:thm4_3}
&{\lambda_l^m}^\star(t)\nonumber\\
=&\frac{f_l^m(\scalemath{0.9}{\mathsf{SINR}}_l^m({\mathbf{p}^m}^\star(t)))}{\scalemath{0.9}{\mathsf{SINR}}_l^m({\mathbf{p}^m}^\star(t))\frac{\displaystyle\partial f_l^m(\scalemath{0.9}{\mathsf{SINR}}_l^m({\mathbf{p}^m}^\star(t)))}{\displaystyle\partial \scalemath{0.9}{\mathsf{SINR}}_l^m({\mathbf{p}^m}^\star(t))}}{p_l^m}^\star(t) {x_l^m}^\star(t).
\end{align}
\item Update $\omega_l^m(t+1)$ with Lagrangian duality-based projection:
\begin{align*}
\omega_l^m(t+1)\leftarrow\frac{\displaystyle r_l^m(t)}{\displaystyle \lambda_l^{m\star}(t)}\Bigg/\sum_{n=1}^M\frac{\displaystyle r_l^n(t)}{\displaystyle\lambda_l^{n\star}(t)}.
\end{align*}
\item Go to Steps 1 to 4 until convergence.
\end{enumerate}
\end{algorithm}


\begin{remark}
Algorithm~\ref{alg:decentralizedAlg} only yields a locally optimal solution to~\eqref{opt:power_mnos_global}.
However, for any $\mathbf{p}^m(0)$ in a sufficiently close neighborhood of ${\mathbf{p}^m}^\star$, the outputs of Algorithm~\ref{alg:decentralizedAlg} converge to the global optimal solutions of~\eqref{opt:power_mnos_global}.
\end{remark}


\section{OpenRANet: An Optimization-based Deep Learning Model with Projection and Convex Optimization Layers for Total Power Minimization}\label{sec:CVXL}

In this section, we design a deep learning model with projection and optimization layers, which we called OpenRANet, to 
obtain an approximate solution to the non-convex problem in \eqref{opt:power_mno_alt}, and then \eqref{opt:power_mnos}. Since \eqref{opt:power_mno_alt} is convex only when a feasible minimum transmission rate requirement of each user $r_l^m$ at each subcarrier is given, the deep learning model seeks to find the appropriate $r_l^m$. We divide our model into four parts: feature extraction with a convolutional filter, a number of dense layers, a projection layer, and a convex optimization layer \cite{boydcvxmodel1,boydcvxmodel2,amos2023tutorial}. We then describe the forward and backward propagation mechanism to tune the learning parameters of our proposed model.

\subsection{The OpenRANet for Total Power Minimization}
\begin{figure*}
\centerline{\includegraphics[scale=0.3]{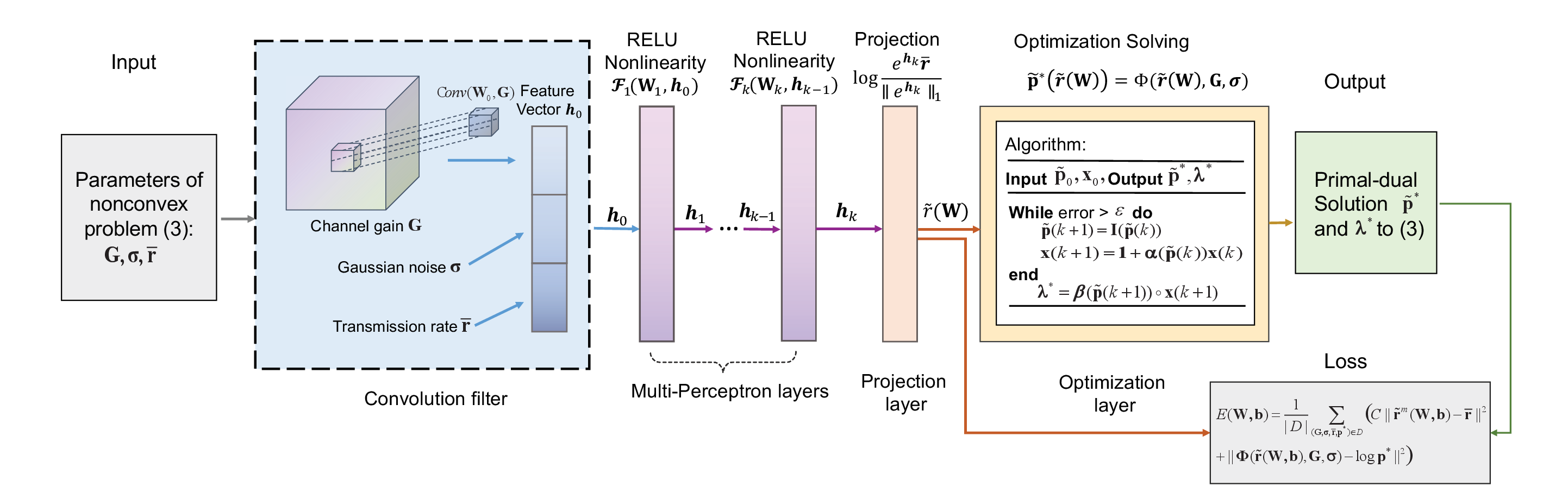}}
\captionsetup{font={footnotesize}, justification=raggedright}
\caption{The architecture of the OpenRANet for approximating the optimal solution to \eqref{opt:power_mno_alt}. The channel gain $G$ is dimensionally reduced by trainable convolutional filters. Then, concatenated with the Gaussian noise $\sigma$ and transmission rate $\bar{\r}$, it is fed into the dense layers as the feature vector. The output $\mathbf{h}_k(\bm{W})$ is then input into the projection layer \eqref{eq:project} such that the transmission rate requirements \eqref{nonconvexity} are satisfied. Finally, the projected transmission rates are input into a convex optimization model, which can be solved using the fast Algorithm \ref{alg:decentralizedAlg}. The loss function considers the deviation of $\bar{\r}$ and $\p^*$, where $C$ is the penalty of violation to \eqref{nonconvexity}.}
\label{fig:wf} 
\end{figure*}

In traditional neural network-based machine learning, we can theoretically fit a complex function with a trained neural network model consisting of linear fully-connected layers and nonlinear activation functions. This is possible if the neural network is given enough capacity and training time, according to the universal approximation theorem \cite{goodfellow2016deep}.
\textcolor{black}{However, for complex optimization problems, particularly the challenging large-scaled non-convex optimization problems, relying solely on machine learning techniques may make it difficult to capture the whole underlying structure of the problems\cite{o2017introduction,song2022benchmarking}. As domain-knowledge techniques have great power in many scenarios, combining modern machine-learning approaches with traditional optimization-based approaches holds significant potential for enhancing effectiveness and efficiency in addressing the complexity challenges of nonconvex optimization \cite{o2017introduction,song2022benchmarking}.}
One way is to identify and extract the convex structure of the complex optimization problem, such as the hidden convex optimization subproblems within it. These convex subproblems can then be integrated into the deep learning model as optimization layers \cite{boydcvxmodel1,boydcvxmodel2,amos2023tutorial,agrawal2020differentiating,agrawal2019differentiable}, resulting in a model that exhibits characteristics of the complex optimization problem. This leads to the deep learning model becoming easier to train, enhancing its generalization ability, and reducing the overall complexity of the model. A brief introduction to the convex optimization layer \cite{boydcvxmodel1,boydcvxmodel2,amos2023tutorial,agrawal2020differentiating,agrawal2019differentiable} is given in  \href{https://arxiv.org/abs/2409.12964} {the ArXiv online version}. Moreover, when dealing with constraints of optimization problems, it is important to incorporate the constraint information into the deep learning model to ensure feasibility. The above considerations make us cautious about relying solely on machine learning methods to solve \eqref{opt:power_mno_alt}. Therefore, we propose OpenRANet, which extensively incorporates the constraint information and the convex subproblem of \eqref{opt:power_mno_alt} into the deep learning model, resulting in significantly improved performance.

Recall from the previous section that when the individual transmission rate requirements $\hat{r}_l^m$ are known, the non-convex problem in \eqref{opt:power_mno_alt} becomes equivalent to the problem in \eqref{opt:power_mno_cvx}. The objective now is to learn the individual transmission rate $r_m^l$ in \eqref{opt:power_mno_alt}. Instead of solely relying on machine learning techniques, we aim to incorporate \eqref{opt:power_mno_cvx} as part of the deep learning model. This means treating \eqref{opt:power_mno_cvx}, which is convex, as a layer within the model to find the solution to the non-convex \eqref{opt:power_mno_alt}. Furthermore, to ensure that the data rate constraints in \eqref{opt:power_mno_alt} are not violated, we first project the output $r_m^l$ onto the feasible domain of problem in \eqref{opt:power_mno_alt} before incorporating \eqref{opt:power_mno_cvx} as an optimization layer into the deep learning model. Consequently, our proposed deep learning model, OpenRANet, whose architecture is shown in Fig. \ref{fig:wf}, is designed with the following components:

\begin{itemize}
    \item Feature extraction: To extract features for a large-scale open RAN system with a high number of users, directly using each system parameter as input features for a deep learning model can result in the curse of dimensionality. To address this issue, convolutional filters can be employed to effectively extract the crucial features of the system as follows:
    \begin{align}\label{eq:conv1}
        &\Tilde{\G}(\bm{W}_0)= \mathcal{Conv}(\bm{W}_0, \G),\nonumber\\
        & \mathbf{h}_0(\bm{W}_0) = [\Tilde{\G}(\bm{W}_0),\bm{\sigma},\bar{\r}]^{\top},
    \end{align}
    where $ \mathcal{Conv}(\cdot)$ represents the convolution operation for the channel gain $\G$ with parameter set $\bm{W}_0$. 
    \item Dense layers: Here we introduce $k$ fully connected layers that incorporate learning parameters $\bm{W}=\{\bm{W}_0,\dots,\bm{W}_0\}$ in the model that can be trained using knowledge of data:
    \begin{align}\label{eq:denseL}
        \mathbf{h}_k(\bm{W})=\mathcal{F}_k(\cdots\mathcal{F}_2(\bm{W}_2(\mathcal{F}_1(\bm{W}_1\mathbf{h}_0(\bm{W}_0)))),
    \end{align}
    where $\mathcal{F}_k(\cdot)$ represents a composite function composed of linear combinations and activation functions such as the RELU function. 
    \item Projection layer: The purpose of this layer is to incorporate the rate constraint information into the OpenRANet, thereby ensuring that the output of the OpenRANet meets the rate requirement. This, in effect, guarantees the practicality of the model. The Projection layer can be designed as:
    \begin{align}\label{eq:project}
        \Tilde{\r}(\bm{W}) = \log \frac{\bm{e}^{\mathbf{h}_k(\bm{W})}\diag(\bar{\r})}{\diag({\bm{e}^{\mathbf{h}_k(\bm{W})}\bm{1}})}.
    \end{align}
    \item Convex optimization layer: This layer integrates the convex optimization subproblem of the nonconvex  \eqref{opt:power_mno_alt} into the model. Note that the log-convexity property in the standard interference function can be used to efficiently solve this convex subproblem, as shown in Theorem \ref{th:dual}. This integration saves significant costs compared to the purely data-driven deep learning model, which would otherwise need to learn all the underlying structures of the \eqref{opt:power_mno_alt}. Consequently, it effectively enhances the performance and efficiency of the model. The optimization layer is designed as:
    \begin{align}\label{eq:conv3}
        \tilde{\p}^{*}(\Tilde{\r}(\bm{W}))=\mathbf{\Phi}(\tilde{\r}(\bm{W}),\G,\bm{\sigma}),
    \end{align}
    where $\mathbf{\Phi}(\tilde{\r}(\bm{W}),\G,\bm{\sigma})$ is the solution to the optimization problem
\begin{align}\label{opt:optL}
    \text{argmin}_{\p}&\ \sum_{l=1}^L\sum_{m=1}^M e^{\tilde{p}_l^m}\nonumber\\
    \text{subject to}&\ \displaystyle \log f_l^m(\scalemath{0.9}{\mathsf{SINR}}_l^m(e^{\tilde{\mathbf{p}}^m}))
    \geq \log \tilde{r}_l^m(\bm{W}), \nonumber\\
    & \quad\forall \;l,\ \forall \; m, \\
    \text{variables:}&\ \tilde{\p}^m=(\tilde{p}_1^m,\cdots,\tilde{p}_L^m)^\top \quad\forall \;m. \nonumber
\end{align}
\end{itemize}

Observe that the output of the projection layer \eqref{eq:project} is the approximate individual transmission rate $\tilde{\r}(\bm{W})$, connecting \eqref{opt:power_mno_alt} and its convex relaxation \eqref{opt:optL}. Actually, \eqref{eq:project} is designed to ensure the transmission rate meets requirement \eqref{nonconvexity}. Also, there is a trade-off coefficient to penalize the violation to \eqref{nonconvexity} in the loss function (\ref{eq:loss}), which is described in the next subsection. The problem in \eqref{opt:optL} can be viewed as a function that maps the system parameters $(\G,\bm{\sigma},\bar{\r})$ and the learning parameters $\bm{W}$ to the solution of \eqref{opt:power_mno_alt}. Given these parameters, the convex relaxation problem in \eqref{opt:optL} can be solved by a fast iterative algorithm, as presented in Algorithm~\ref{alg:decentralizedAlg}. Note that \eqref{eq:iter_p} in Algorithm~\ref{alg:decentralizedAlg} is the iteration of the standard interference function of transmission rate function $f_l^m(\scalemath{0.9}{\mathsf{SINR}}_l^m({\mathbf{p}^m}(k)))$ in the constraints (cf. Lemma \ref{p_standard}), while \eqref{eq:virtual_x} is due to the stationarity of the Lagrangian of \eqref{opt:optL}. Also, the Algorithm~\ref{alg:decentralizedAlg} not only gives the primal solution $\p^*$ but the dual solution $\bm{\lambda}^*$ to the problem in \eqref{opt:optL}, as we can obtain $\lambda^*$ by \eqref{eq:thm4_3} using the output $\p^*$ and $\x^*$. This is important as the dual can also be used in the backward propagation to tune the learning parameters, as we will show in the following section. 
\textcolor{black}{
Suppose the output of the convolution filter (one can choose an appropriate size for the kernel) is $K$. Then the dimension of $\mathbf{h}_0(\bm{W}_0)$ in \eqref{eq:conv1} is $(2ML+K)\times 1$. The dense layers are used for the approximation of the optimal data rate $r_{l}^{m*}$ per user per subcarrier; thus its dimension is $M\times L$. The output \eqref{eq:denseL} is then input to the projection layer in \eqref{eq:project}, whose output dimension is still $M\times L$. The output of the projection layer in \eqref{eq:project} is then used to update $\tilde{r}(\bm{W})$ in the convex optimization layer in \eqref{opt:optL}, and the output of the \eqref{opt:optL} is the optimal solution of $\mathbf{p}$, whose dimension is also $M\times L$. 
}

Combining \eqref{eq:conv1}-\eqref{eq:conv3} and the  Algorithm~\ref{alg:decentralizedAlg} for solving \eqref{eq:conv3}, the forward pass stage of our proposed OpenRANet is presented in ``Forward Propagation" in Algorithm~\ref{alg:cvxDL}. From the perspective of end-to-end learning, the learning parameters $\bm{W}$ are trained to approximate the optimal solution to \eqref{opt:power_mno_alt}. Compared to other traditional data-driven deep learning models (e.g., one can simply train a CNN only to output the $r_m^l$ in \eqref{opt:power_mno_cvx} and use this output to solve \eqref{opt:optL} and then \eqref{opt:power_mno_alt}), tuning these learning parameters not only depend on a universally applicable black-box-like convolutional filter and fully-connected neural network, but also the projection layer \eqref{eq:project} and the convex optimization layer \eqref{opt:optL}, which capture the specific structure and information of \eqref{opt:power_mno_alt}. This significantly improves both the accuracy and interpretability of the model. Also, it reduces the number of learning parameters in our model compared to the pure data-driven learning models. We will present these benefits in the simulation section.

\subsection{Training Process}
To train the learning parameters $\bm{W}$ in the OpenRANet designed above for the non-convex problem in \eqref{opt:power_mno_alt}, we need forward and backward propagation for our model. Suppose we are given a dataset $\bm{D}$ with $|\bm{D}|$ training samples. The loss function between the output of our learning model and the ground truths is defined as
\begin{align}
\label{eq:loss}
   E(\bm{W}) = &\frac{1}{|\bm{D}|} \sum\limits_{(\G,\bm{\sigma},\bar{\r},\p^*)\in \bm{D}}\big(C||\sum_{m=1}^M\Tilde{\r}^m(\bm{W})-\bar{\r}||^2\nonumber\\
    &+||\mathbf{\Phi}(\tilde{\r}(\bm{W}),\G,\bm{\sigma})-{\p}^*||^2\big),
\end{align} 
where $C$ is a trade-off of the violation of the constraint $\sum_{m=1}^M e^{\tilde{r}_l^m} \geq \bar{r}_l$, e.g., if we cannot tolerate any violation of this constraint, we can set $C$ to a large positive number and vice versa. 

The dataset $ \bm{D} $ in \eqref{eq:loss} includes labeled training data with channel gain, noise power, and the minimal required transmission rate as inputs, while the optimal solution $ \p^* $ serves as the output of the OpenRANet model. In practical applications, this dataset can be obtained through various methods, such as simulations or historical data. To create the dataset, we can first conduct simulations of the Open RAN environment by modeling different scenarios that encompass various user distributions, channel gains, noise power levels, and required transmission rates. In these simulations, we compute the optimal power allocation for each scenario using established algorithms, like the branch and bound algorithm \cite{zhai2017rate, balakrishnan1991branch}, which systematically explores the solution space to find optimal allocations while meeting constraints. Additionally, we can gather historical data from existing Open RAN deployments, which may include real-world measurements of channel conditions, noise levels, and actual power allocations used to meet data rate requirements. This historical data helps validate and refine the synthetic dataset, ensuring it reflects realistic operating conditions and variations. By combining these two sources, we can create a robust training dataset that captures a wide range of scenarios, enhancing the model's ability to generalize and accurately predict optimal power allocations in real applications.

To compute the gradient of the loss for the learning parameters $\bm{W}$ in the backward propagation, we first need to derive $\frac{\partial \mathbf{\Phi}(\tilde{\r}(\bm{W}),\G,\bm{\sigma})}{\partial \tilde{\r}(\bm{W})}$ for the last convex optimization layer, and then derive $\frac{\partial \tilde{\r}(\bm{W})}{\partial\bm{W}}$ for the previous layers of the CNN in our model. In what follows, we will present the derivation of $\frac{\partial \mathbf{\Phi}(\tilde{\r}(\bm{W}),\G,\bm{\sigma})}{\partial \tilde{\r}(\bm{W})}$ briefly. Given the input  $\tilde{\r}(\bm{W})$, the problem in \eqref{opt:optL} is convex and its primal-dual solution  $(\tilde{\p}^*,\bm{\lambda}^*)$ can be effectively obtained by using Algorithm \ref{alg:decentralizedAlg}. Therefore, we can consider the primal-dual solution as a function of  $\tilde{\r}(\bm{W})$, i.e., $\tilde{\p}^*(\tilde{\r}(\bm{W}))$ and $\bm{\lambda}^*(\tilde{\r}(\bm{W}))$.
The key aspect to differentiating $\tilde{\p}^*(\tilde{\r}(\bm{W}))$ for $\tilde{\r}(\bm{W})$ is to differentiating the KKT conditions of \eqref{opt:optL} by using matrix calculus techniques, as the solution of \eqref{opt:optL} can be found by finding the root of the KKT conditions. Specifically, let us denote $\bm{\mathcal{K}}(\tilde{\p},\bm{\lambda})$ as
\begin{align*}
\bm{\mathcal{K}}(\tilde{\p},\bm{\lambda},\tilde{\r}) \!\!=\!\!\! \begin{bmatrix}\!\!
\nabla_{\tilde{\p}}(\sum\limits_{m}\sum\limits_{l}e^{\tilde{p}_l^m}\!)\!+\!\partial_{\tilde{\p}}(\sum\limits_{m}\sum\limits_{l}\lambda_l^m\log\frac{r_l^m}{f_l^m(\mathsf{SINR(e^{\tilde{p}_l^m})})})\!\!\\ 
\bm{\lambda} \circ (\log(\tilde{\r})-\log\bm{f}(\mathsf{SINR(e^{\tilde{p}})}))
\end{bmatrix}.
\end{align*}
Then the KKT conditions of \eqref{opt:optL} are
$\bm{\mathcal{K}}(\tilde{\p}^*(\tilde{\r}),\bm{\lambda}^*(\tilde{\r}),\tilde{\r})=0,\;\log\bm{f}(\mathsf{SINR(e^{\tilde{p}^*})})\geq\log(\tilde{\r}),$ and $\bm{\lambda}^*(\tilde{\r})\geq \bm{0}$.
Since we have $\log\bm{f}(\mathsf{SINR(e^{\tilde{p}^*})})= \tilde{\r}$ from Theorem \ref{theorem2}, finding the solution of \eqref{opt:optL} is simply finding the root of
\begin{align}\label{eq:kkt_con}
   \bm{\mathcal{K}}(\tilde{\p}^*(\tilde{\r}),\bm{\lambda}^*(\tilde{\r}),\tilde{\r})=0. 
\end{align}
As the equation set \eqref{eq:kkt_con} always holds for any $\tilde{\r}$ with which the problem in \eqref{opt:optL} has feasible solutions, we have
$$\frac{\partial}{\partial\tilde{\r}}\bm{\mathcal{K}}(\tilde{\p}^*(\tilde{\r}),\bm{\lambda}^*(\tilde{\r}),\tilde{\r})=\bm{0}.$$
Using the chain rule for the above derivation, we have
\begin{align}\label{eq:deriv_equa}
   \frac{\partial (\tilde{\p}^*(\tilde{\r}),\bm{\lambda}^*(\tilde{\r}))}{\partial\tilde{\r}}= -\big(\frac{\partial \bm{\mathcal{K}}(\tilde{\p}^*,\bm{\lambda}^*,\tilde{\r})}{\partial(\tilde{\p}^*,\bm{\lambda}^*)}\big)^{-1} \frac{\partial\bm{\mathcal{K}}(\tilde{\p}^*,\bm{\lambda}^*,\tilde{\r})}{\partial \tilde{\r}},
\end{align}
where
\begin{align}\label{eq:hessian}
&\frac{\partial \bm{\mathcal{K}}(\tilde{\p}^*,\bm{\lambda}^*,\tilde{\r})}{\partial(\tilde{\p}^*,\bm{\lambda}^*)}\nonumber\\
&=
\begin{bmatrix}
\nabla^2_{\tilde{\p}}(\sum\limits_{m}\sum\limits_{l}e^{\tilde{p}_l^{m*}})\\+\partial^2_{\tilde{\p}}(\sum\limits_{m}\sum\limits_{l}\lambda_l^{m*}\log\frac{r_l^m}{f_l^m(\mathsf{SINR(e^{\tilde{p}_l^{m*}})})}) & \partial^2_{\tilde{\p}}\log\frac{r}{\bm{f}(\mathsf{SINR(e^{\tilde{p}^{*}})}}\\ 
 \partial^2_{\tilde{\p}}\log\frac{\tilde{\r}}{f_l^m(\mathsf{SINR(e^{\tilde{p}^{m}})}}\bm{\lambda^*}& \log\frac{\tilde{\r}}{\bm{f}(\mathsf{SINR(e^{\tilde{p}^{*}})}}
\end{bmatrix}.
\end{align}
Combining \eqref{eq:deriv_equa} and \eqref{eq:hessian}, we can compute the derivatives $ \frac{\partial (\tilde{\p}^*(\tilde{\r}),\bm{\lambda}^*(\tilde{\r}))}{\partial\tilde{\r}}$ in \eqref{eq:deriv_equa} when given the primal-dual solution to \eqref{opt:optL} (computed via Algorithm~\ref{alg:decentralizedAlg}). Then we can further obtain  $\frac{\partial \mathbf{\Phi}(\tilde{\r}(\bm{W}),\G,\bm{\sigma})}{\partial \tilde{\r}(\bm{W})}$.

Deriving $\frac{\partial \mathbf{\Phi}(\tilde{\r}(\bm{W}),\G,\bm{\sigma})}{\partial \tilde{\r}(\bm{W})}$ using  \eqref{eq:deriv_equa} and \eqref{eq:hessian} is actually a general approach. For some specific problems, there are sometimes more convenient ways to obtain this derivative. Note that the input $\tilde{\r}(\bm{W})$ to \eqref{opt:optL} is given in the constraints, which become tight when \eqref{opt:optL} is at optimality, so that we can also derive $\frac{\partial \mathbf{\Phi}(\tilde{\r}(\bm{W}),\G,\bm{\sigma})}{\partial \tilde{\r}(\bm{W})}$ by considering the tight constraints. For example, for the transmission rate function defined as the Shannon Capacity in TABLE \ref{table:ratefunction}, this derivative can be obtained by using the tight constraints as
\begin{align}\label{eq:explicitgradienty1}
    \frac{\partial\tilde{p}_i^{m\star}}{\partial\tilde{r}_j^m}= \frac{r_j^m}{p_i^{m\star}}\frac{\partial\mathbf{e}_i^T\mathbf{p}^{m\star}}{\partial\mathbf{R}_{jj}^m}e^{r_j^m}.
\end{align}
More details of the derivation are available in \href{https://arxiv.org/abs/2409.12964} {the ArXiv online version}.

 Next, we present the backward pass stage of the deep learning model in the ``Backward Propagation" in Algorithm~\ref{alg:cvxDL}.

\begin{algorithm}
\caption{The OpenRANet.}
\label{alg:cvxDL}
\SetAlgoLined
 \KwIn{Problem instances with the channel gain $\G$, noise power $\bm{\sigma}$ and $\bar{\r}$ in training dataset $D$.} 
 \KwOut{Power allocation $\tilde{\p}$*.}
\textbf{$\quad$Forward propagation:}\\
\begin{enumerate}
\item Extract the features of \eqref{opt:power_mno_alt} by using a convolutional filter:
\begin{align*}
        &\Tilde{\G}(\bm{W}_0)= \mathcal{Conv}(\bm{W}_0, \G),\nonumber\\
        & \mathbf{h}_0(\bm{W}_0) = [\Tilde{\G}(\bm{W}_0),\bm{\sigma},\bar{\r}]^{\top}.
    \end{align*}
\item Obtain the approximate $\tilde{\r}(\bm{W})$ the dense layers and projection layer:
\begin{align*}
    &\mathbf{h}_k(\bm{W})=\mathcal{F}_k(\cdots\mathcal{F}_2(\bm{W}_2(\mathcal{F}_1(\bm{W}_1[\Tilde{\G}(\bm{W}_0),\bm{\sigma},\bar{\r}]^{\top}))),\\
&\Tilde{\r}(\bm{W}) = \log \frac{\bm{e}^{\mathbf{h}_k(\bm{W})}\diag(\bar{\r})}{\diag({\bm{e}^{\mathbf{h}_k(\bm{W})}\bm{1}})}.
\end{align*}
\item Input $\tilde{\r}(\bm{W})$ into the convex optimization layer $\mathbf{\Phi}(\tilde{\r}(\bm{W}),\G,\bm{\sigma})$ defined as  \eqref{opt:optL} and use  Algorithm~\ref{alg:decentralizedAlg} to obtain the primal-dual optimal solutions $\tilde{\p}^*(\tilde{\r})$ and $\tilde{\lambda}^*(\tilde{\r})$.

$\!\!\!\!\!\!\!\!\!$\textbf{Backward Propagation:}
\item Input the $\tilde{\r}(\bm{W})$ and $\tilde{\p}^*(\tilde{\r})$ into the loss function in (\ref{eq:loss}) with its derivative $\partial E(\bm{W})/\partial \bm{W}$ with respect to the hyperparameter $\bm{W}$ as:
\begin{align*}\label{eq:deri_loss}
&\frac{1}{|D|}\!\! \sum\limits_{(\G,\bm{\sigma},\bar{\r},\p^*)\in \bm{D}}\!\!\!\!\!\!\!\!\!\!\Big(2C(\tilde{\r}(\bm{W})\!-\!\bar{\r})\!\!^{\top}\!\frac{\partial \Tilde{\r}(\bm{W})}{\partial \bm{W}} 
    \nonumber\\
    +& 2\big(\mathbf{\Phi}(\tilde{\r}(\bm{W}),\!\G,\bm{\sigma})\!\!-\!\!\tilde{\p}^*\big)^{\top}
    \!\!\!\times \!\!\frac{\partial \mathbf{\Phi}(\tilde{\r}(\bm{W}),\!\G,\bm{\sigma})}{\partial \tilde{\r}(\bm{W})} \frac{\partial \Tilde{\r}(\bm{W})}{\partial \bm{W}}\Big)\!,
\end{align*}
where $\frac{\partial \mathbf{\Phi}(\tilde{\r}(\bm{W}),\G,\bm{\sigma})}{\partial \tilde{\r}(\bm{W})}$ can be obtained by \eqref{eq:deriv_equa} and \eqref{eq:hessian} with $\tilde{\p}^*(\tilde{\r})$ and $\tilde{\lambda}^*(\tilde{\r})$.
\item Update the learning parameter $(\bm{W})$ by
$\bm{W} \leftarrow \bm{W} - \alpha \frac{\partial E(\bm{W})}{\partial \bm{W}}$,
where $\alpha$ is the learning rate.
\end{enumerate}
\end{algorithm}

\textcolor{black}{The additional complexity introduced by OpenRANet manifests in the following ways:
During the forward propagation phase, the computational complexity required by the projection layer is not significant. Its primary function is to project the output of the dense layers onto the feasible set to satisfy the rate requirement, with computational complexity at $O(ML)$. The main increase in complexity, compared to purely machine learning models, resides in achieving the optimal solutions for the convex optimization layer problem. However, this complexity can be effectively managed through the iterative method outlined in Theorem 2. These iterations converge geometrically in accordance with Krause's theorem, as referenced in \cite{krause2001concave}.
During the backward propagation phase, the most significant increase in complexity for OpenRANet stems from reversing the KKT conditions, which is necessary to update the learning parameters.
Nonetheless, OpenRANet's complexity can be mitigated in certain aspects as delineated below:
The integration of domain-specific knowledge (i.e., the convexity of the subproblems) into the OpenRANet effectively streamlines both the design and training phases of the model. This streamlining leads to reductions in training costs in two main ways: 1) Eliminating the need for additional learning layers and learning parameters to capture the entire underlying structure of the problem alleviates the complexities involved in the backward propagation phase. By decreasing the number of learning layers and parameters, we can reduce the application of hyperparameters, which in turn reduces the number of attempts needed to design different models and then diminishes both the difficulty of training and the associated costs.
Overall, the computational complexity of the proposed OpenRANet might not always be much lower than the pure machine learning models, but the OpenRANet has the potential to achieve more precise or theoretically sound solutions. The choice between these approaches should consider the specific requirements of the task, including the trade-off between computational resources and model performance.}

\begin{remark}
The OpenRANet algorithm outperforms in adhering to problem constraints, solution accuracy, and computational efficiency, surpassing pure machine learning and optimization-based methods, as demonstrated in Section \ref{sec:experiments}.
    The implementation of the learning-based OpenRANet model in Open RAN involves collecting data from various network elements, allowing the Non-Real-Time RAN Intelligent Controller (RIC) to train the OpenRANet and the Near-Real-Time RIC to deploy it for real-time decision-making and performance optimization. 
Some of the key interfaces in this process include O1, A1, and E2. The O1 interface, at the management layer, gathers historical data from network elements, creating a robust training dataset. Following data collection, feature engineering extracts relevant features like channel gains, noise power, and transmission rates for training the OpenRANet.
The A1 interface connects the Non-Real-Time RIC and Near-Real-Time RIC, facilitating the integration of the OpenRANet into network management by exchanging optimization policies for effective power allocation. The E2 interface, between the RAN and Near-Real-Time RIC, enables real-time data collection, allowing the trained OpenRANet to make immediate decisions based on live data. By leveraging these interfaces, the OpenRANet achieves comprehensive power optimization during both training and deployment, ensuring efficient network operation. 
\end{remark}

\begin{remark}
    Despite the effectiveness of the OpenRANet, pre-training it once and deploying it directly to Open RAN might result in performance decline due to evolving traffic patterns, including fluctuations in network conditions, user density changes, temporal variations, and sudden spikes. To create a robust framework for maintaining and enhancing the performance of the OpenRANet in dynamic environments, it is crucial to integrate training techniques such as transfer learning, incremental learning, scheduled retraining, trigger-based retraining, and so on. These techniques are interconnected and work synergistically with the OpenRANet model.  Due to space constraints, this paper primarily focuses on the model mechanism for our proposed nonconvex power minimization problem. However, the current framework of OpenRANet can serve as a foundation for extensions to accommodate varying traffic patterns. In future work, we can consider adapting the model to handle varying traffic patterns. We believe that this is an important concern that deserves more in-depth investigation.
\end{remark}

\section{Numerical Examples}\label{sec:experiments}

In this section, we first provide some simulation results on the performance of the reweighted primal-dual algorithm described in Algorithm~\ref{alg:decentralizedAlg} for local optimality of the problem in \eqref{opt:power_mno_alt} with the transmission rate function for CDMA approximation. Then we demonstrate the performance of the proposed OpenRANet to approximate optimal solutions to  \eqref{opt:power_mno_alt} under different transmission rate constraints. The data and code to generate random samples, implement the primal-dual algorithm, and implement the deep learning models to replicate the numerical experiments are at \url{https://github.com/convexsoft/OpenRANet}.

\subsection{Performance of the Primal-dual Algorithm}\label{sec:eval1}

Consider an open RAN system having $4$ users communicating by way of $2$ subcarriers. We simulate the wireless channel under Rayleigh, Rician, Nakagami, and Weibull fading assumptions. Let $G_{ij}$ stand for the non-negative path gain and $d_{ij}$ stand for the Euclidean distance between the $i$-th user and the $j$-th user. We adopt the well-known path loss model~\cite{3GPP} $G_{ij}^m=10^{-12.8}d_{ij}^{-3.76}$, where $10^{-12.8}$ is the attenuation factor that represents power variations due to path loss. Using $A_{ij}$ to model the signal amplitude (not power) under specific fading types, we represent the interference power received at the $j$-th receiver from the $i$-th user by $G_{ij}A_{ij}^2p_j$. Typically, an approximate $25$dB \textsf{SINR} value is recommended for medium- to high-quality data networks, so we rescale the diagonal entries in $G$ by approximately $10^{25/10}$ times $\mathsf{num\_ues}$ to compensate for extra techniques, such as antenna gain and coding gain. We synthesize the channel gain data with the \textsf{MATLAB}\textsuperscript{\textregistered} Propagation and Channel Model toolbox. In addition, we remove the transmit power upper bound constraints for tractability and assume the same noise level $\sigma_j^m=-60$dBm applies for all $j$ and $m$. The transmission rate requirements are generated with the typical daily relative traffic load pattern described in \cite{10.1109/MWC.2011.6056690}. All the parameters are chosen to make~\eqref{opt:power_mno_alt} feasible.


\begin{figure}[htbp]
\centering
\vspace{-6mm}
\subfigure[ ]{
\begin{minipage}[t]{0.45\linewidth}
\centering
\includegraphics[width=1.6in]{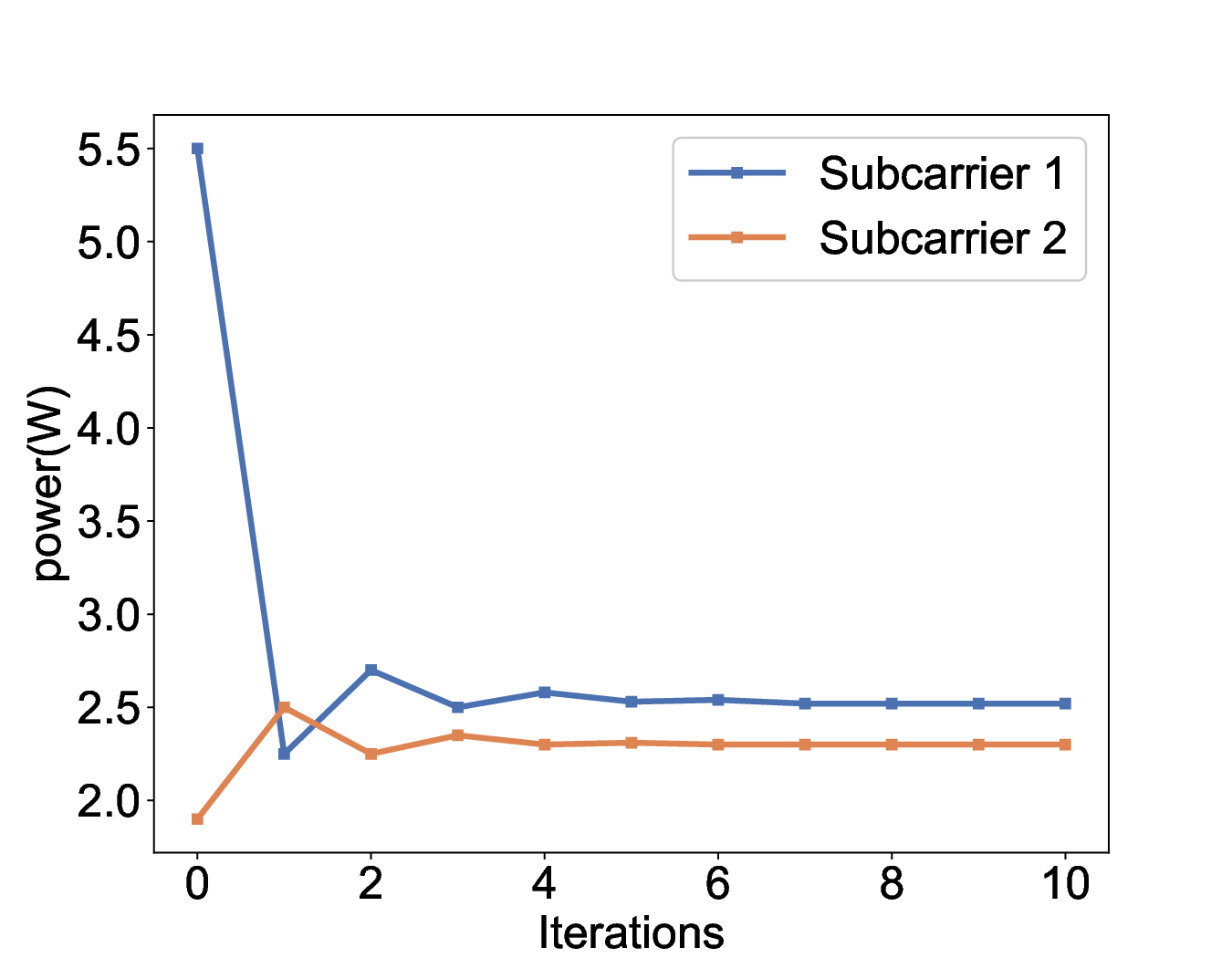}
\end{minipage}%
}%
\subfigure[ ]{
\begin{minipage}[t]{0.45\linewidth}
\centering
\includegraphics[width=1.6in]{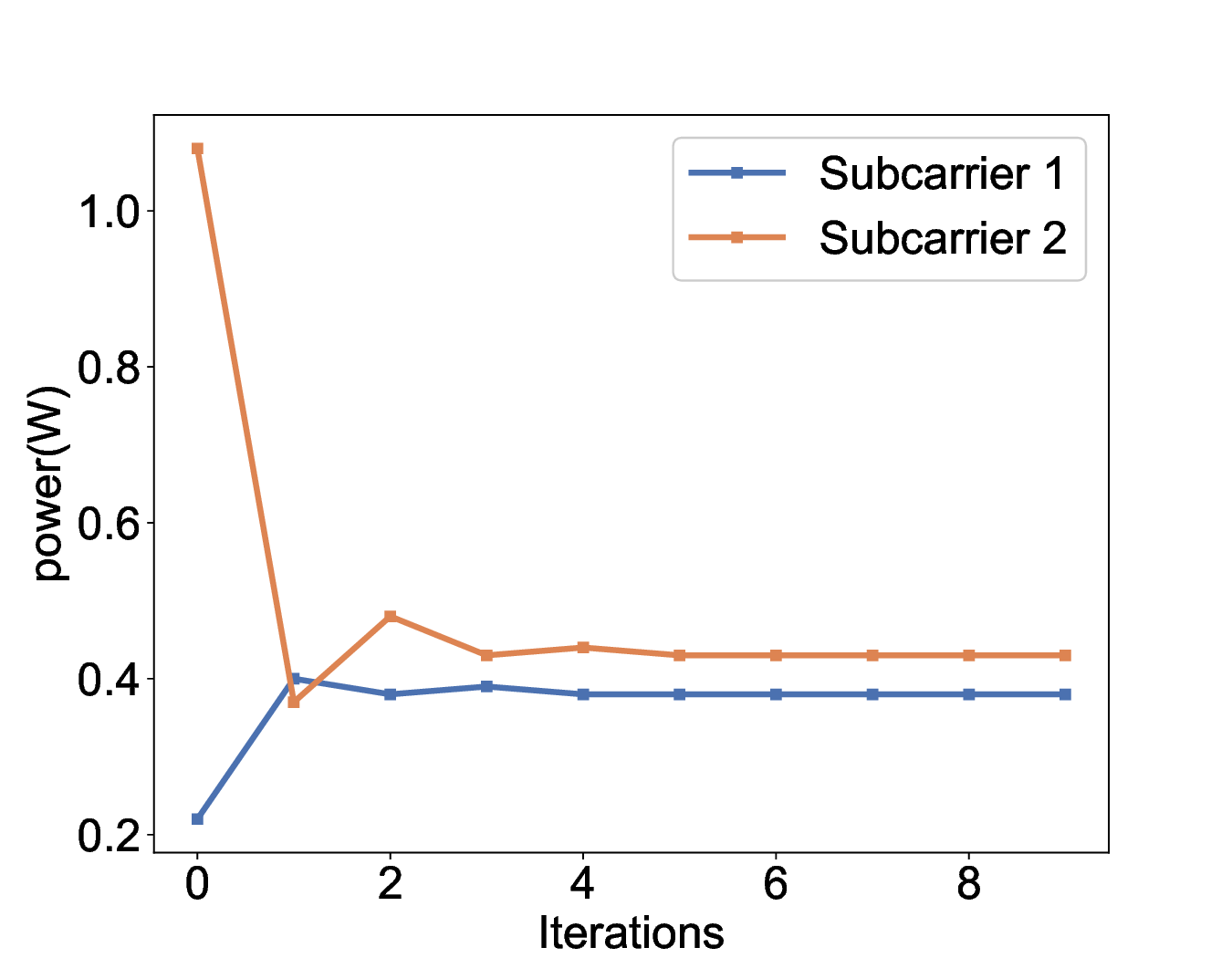}
\end{minipage}%
}%
\vspace{-3mm}
\subfigure[ ]{
\begin{minipage}[t]{0.45\linewidth}
\centering
\includegraphics[width=1.6in]{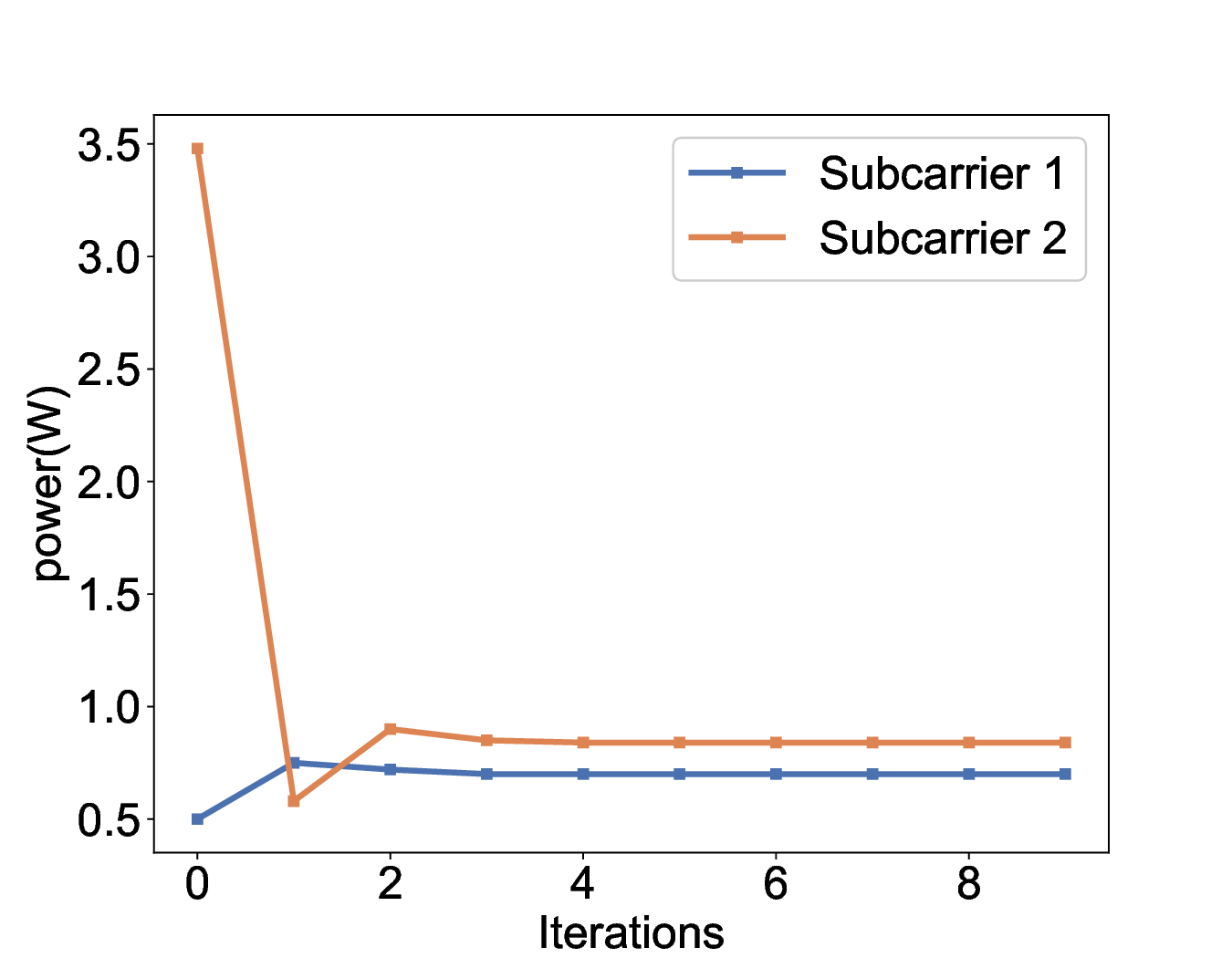}
\end{minipage}%
}%
\subfigure[ ]{
\begin{minipage}[t]{0.45\linewidth}
\centering
\includegraphics[width=1.6in]{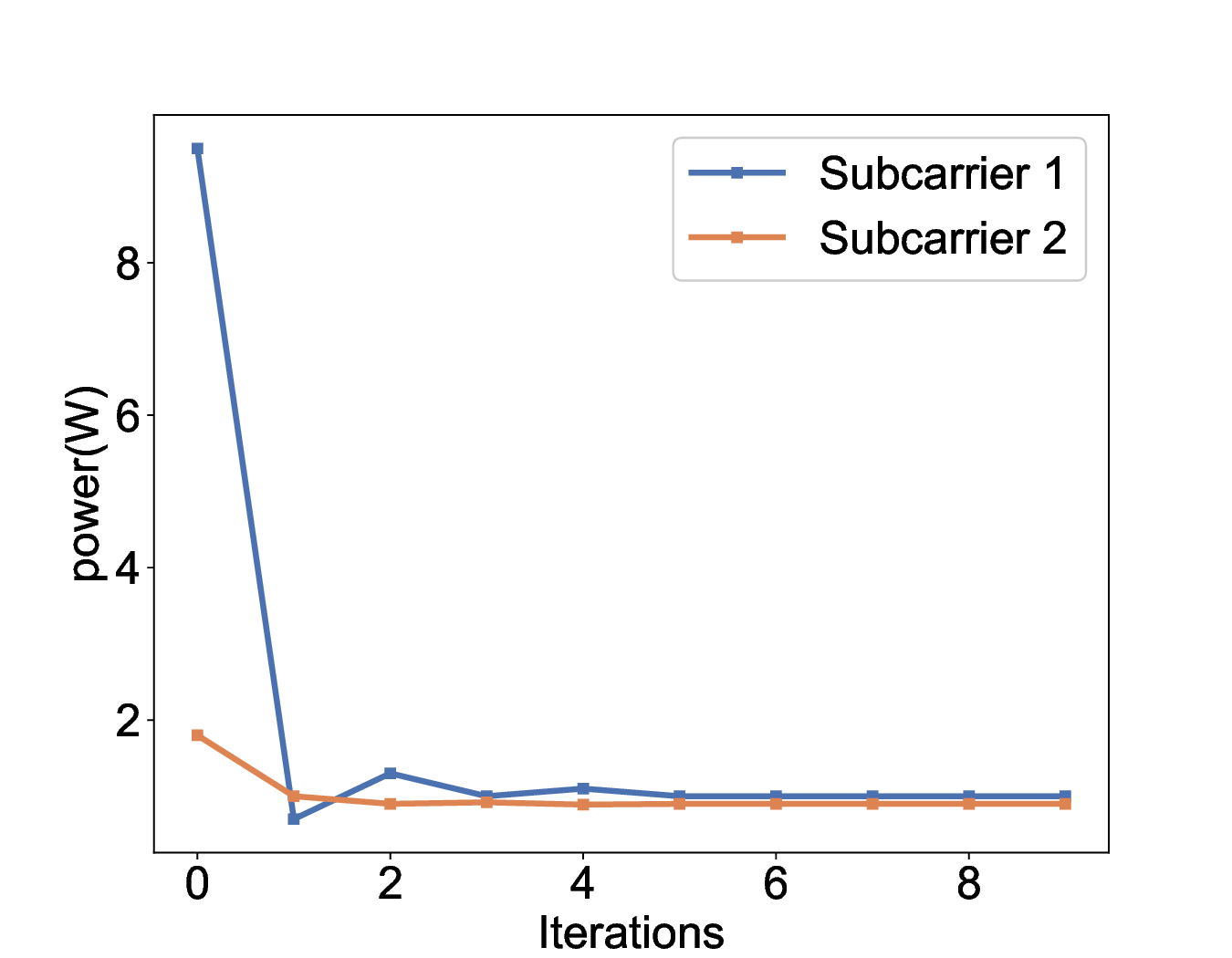}
\end{minipage}%
}%
\vspace{-2mm}
\centering
\captionsetup{font={footnotesize}, justification=raggedright}
\caption{Power allocation per subcarrier with Algorithm~\ref{alg:decentralizedAlg} for the 4-users by 2-subcarriers wireless network, under Rayleigh (a), Rician (b), Nakagami (c), and Weibull fading (d) channels, resp. The 95\% confidence intervals were plotted for 100 trials over the same random sample.}
\label{fig:ap_2_4}
\end{figure}

\begin{figure}[htbp]
\centering
\vspace{-6mm}
\subfigure[ ]{
\begin{minipage}[t]{0.45\linewidth}
\centering
\includegraphics[width=1.6in]{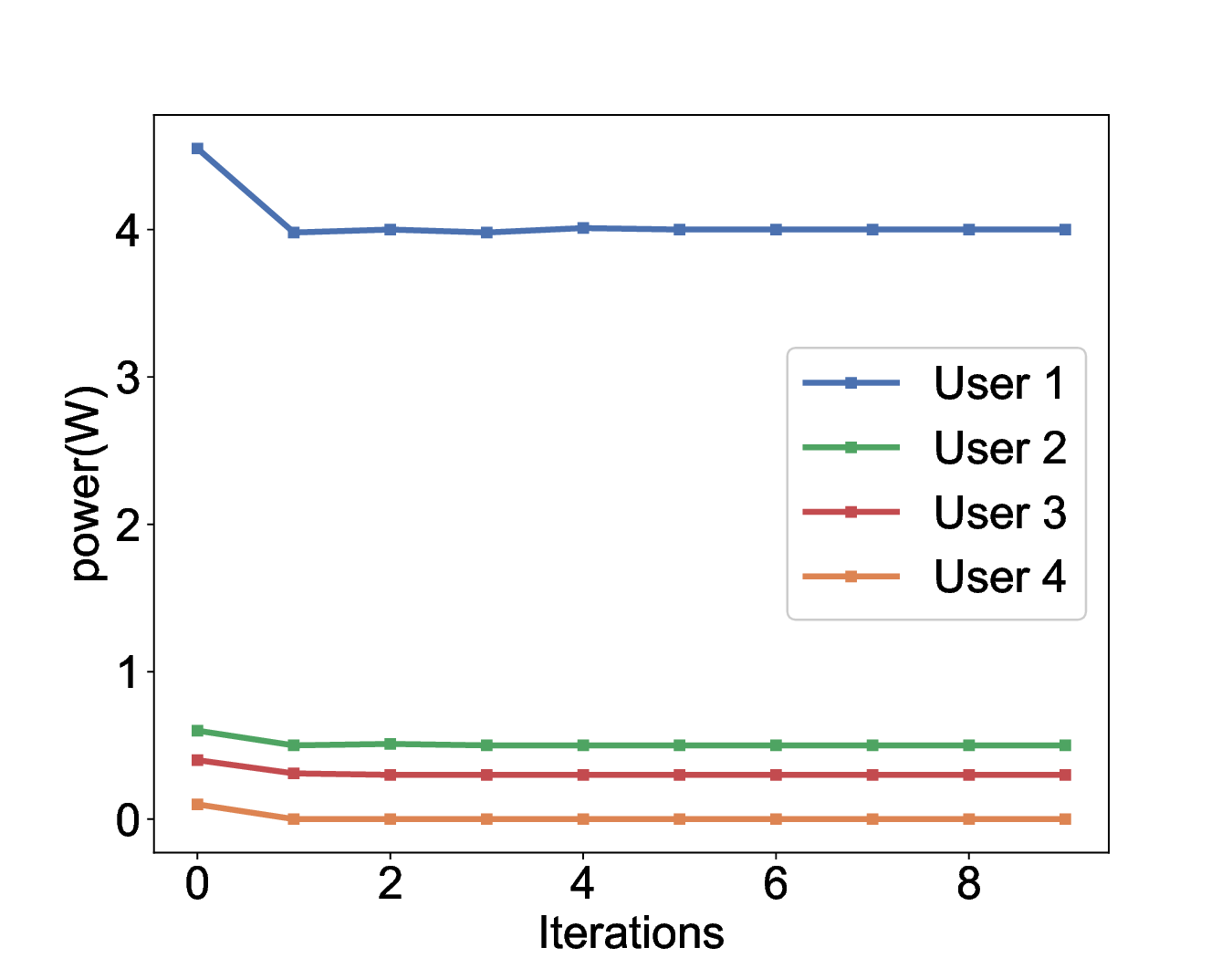}
\end{minipage}%
}%
\subfigure[ ]{
\begin{minipage}[t]{0.45\linewidth}
\centering
\includegraphics[width=1.6in]{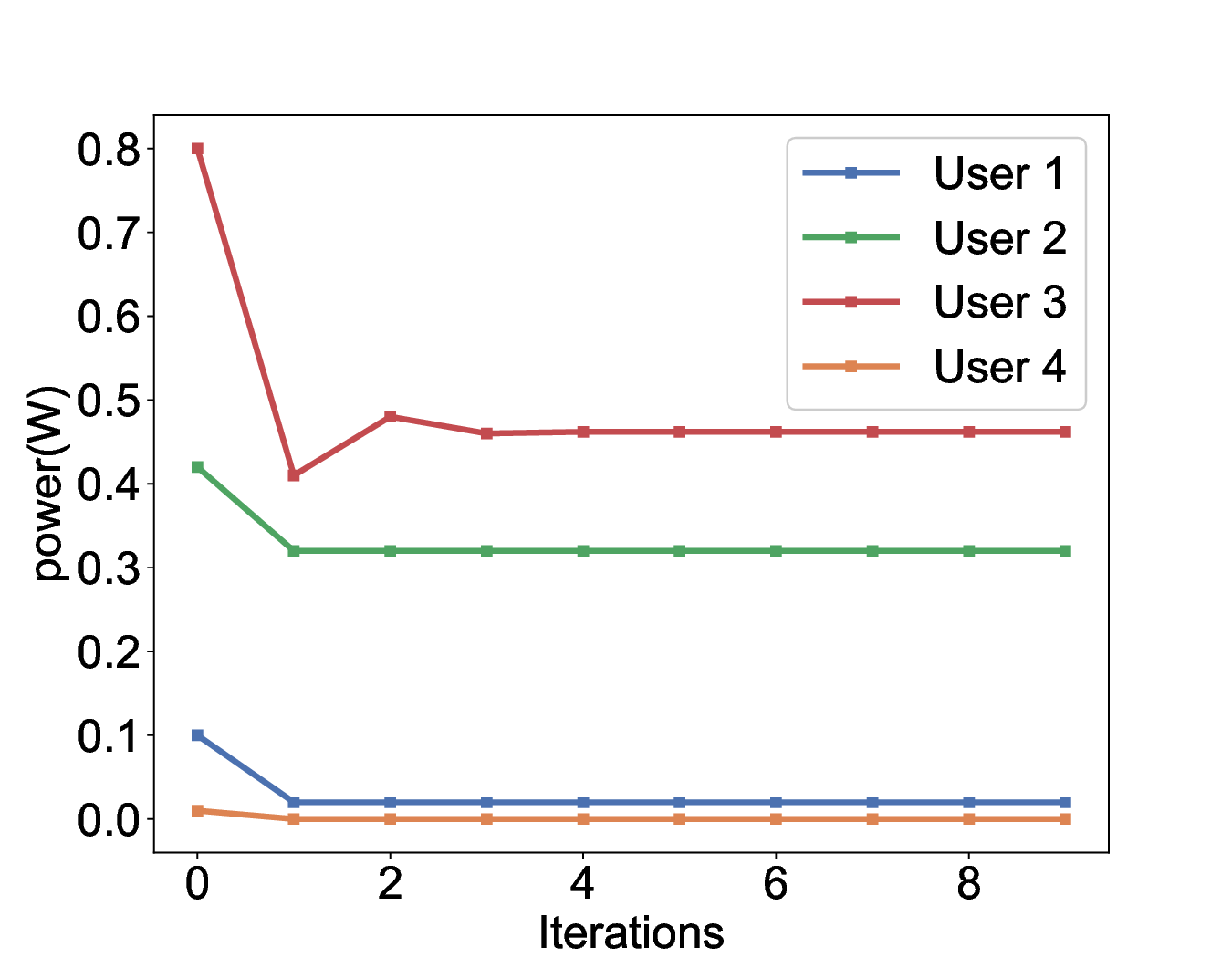}
\end{minipage}%
}%
\vspace{-3mm}
\subfigure[ ]{
\begin{minipage}[t]{0.45\linewidth}
\centering
\includegraphics[width=1.6in]{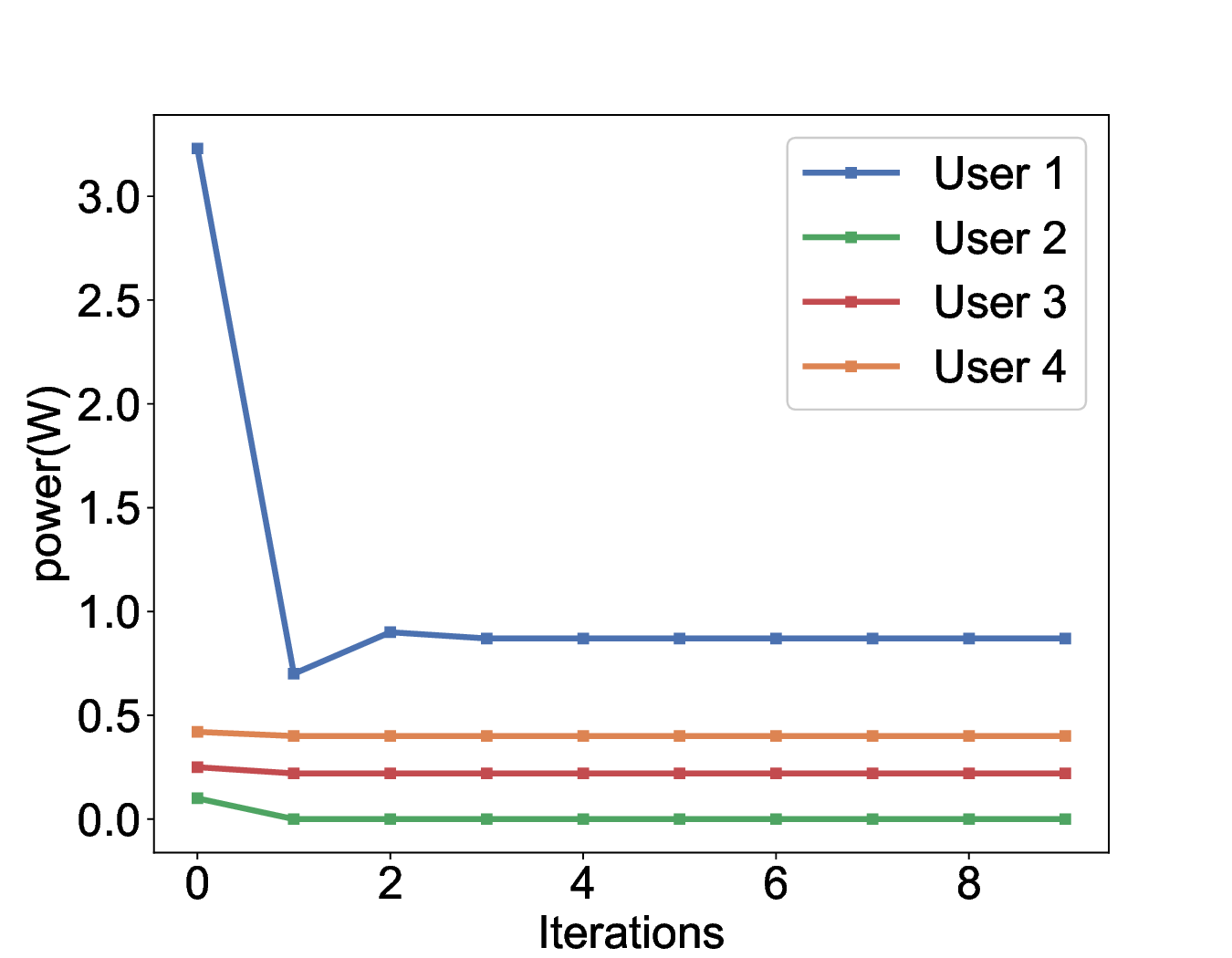}
\end{minipage}%
}%
\subfigure[ ]{
\begin{minipage}[t]{0.45\linewidth}
\centering
\includegraphics[width=1.6in]{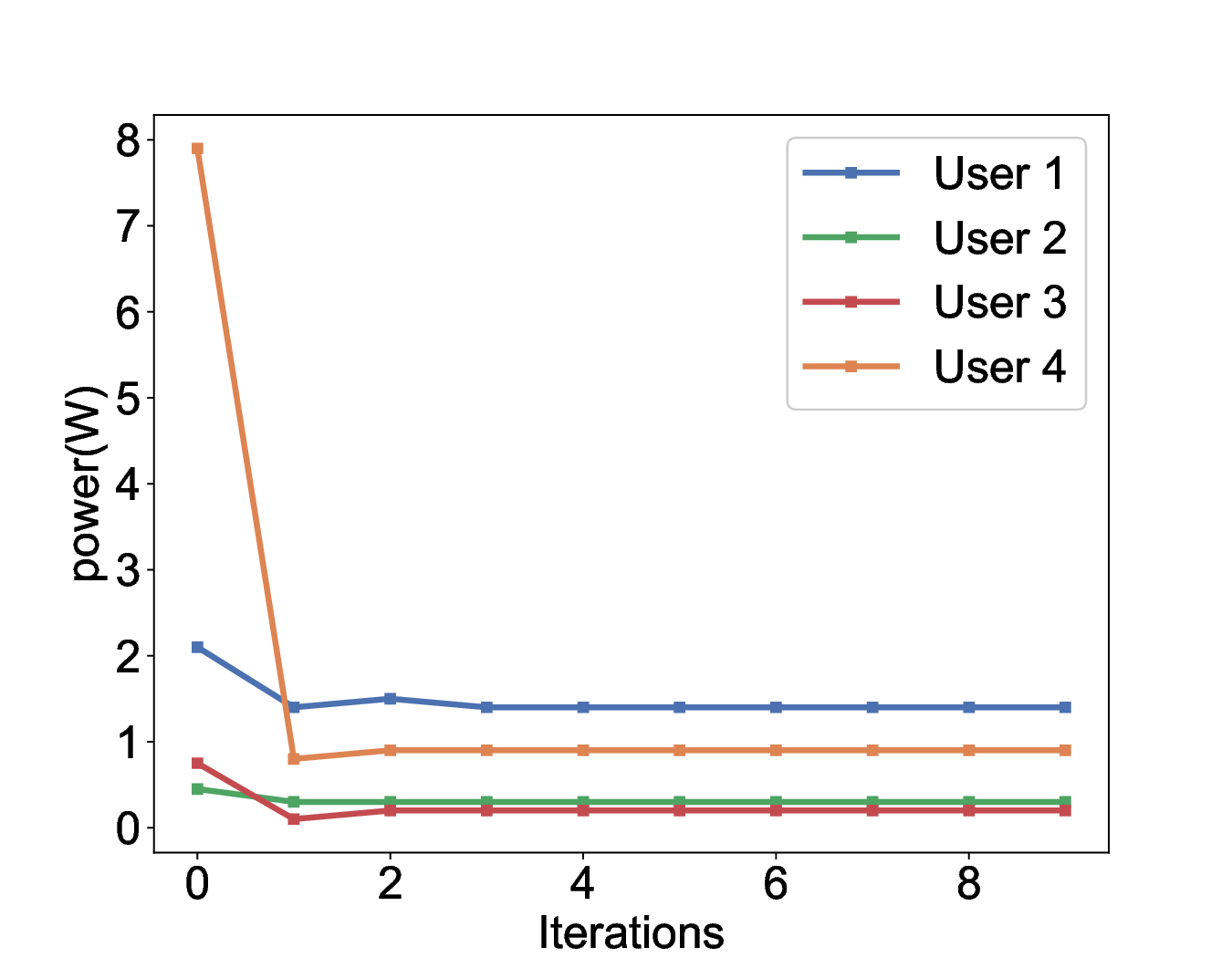}
\end{minipage}%
}%
\centering
\captionsetup{font={footnotesize}, justification=raggedright}
\caption{Power allocation per user with Algorithm~\ref{alg:decentralizedAlg} for the 4-users by 2-subcarriers wireless network, under Rayleigh (a), Rician (b), Nakagami (c), and Weibull fading (d) channels, resp. The 95\% confidence intervals were plotted for 100 trials over the same random sample.}
\label{fig:ue_2_4}
\end{figure}


Fig.~\ref{fig:ap_2_4} and~\ref{fig:ue_2_4} illustrate the convergence of Algorithm~\ref{alg:decentralizedAlg} and the evolution of the individual transmit powers. Both figures show the convergence to an equilibrium solution, which corresponds to a local optimality. It is observed that Algorithm~\ref{alg:decentralizedAlg} almost always converges to the same solution, and the random initialization of the weights (and hence $r_l^m$) does not have much impact on optimality. Besides, the $95\%$ confidence intervals nearly vanish after the first few iterations, implying a desirable fast-converging performance. Besides these numerical examples with 2 subcarriers and 4 users, similar numerical behaviors have been observed in networks with tens of users and subcarriers.




\subsection{Performance of the OpenRANet}\label{sec:eval2}

As shown in Fig. \ref{fig:wf}, the OpenRANet uses some dense layers to approximate the transmission rate, a projection layer to ensure the feasibility of the solution of \eqref{opt:power_mno_alt}, and a final convex optimization layer to obtain the solution of \eqref{opt:power_mno_alt}. The inputs to the model are the magnitude of the channel coefficient $\G^m$, the transmission rate requirement $\bar{\r}$, and the additive Gaussian noise sigma $\bm{\sigma}^m$. The outputs of the model are the transmission rate $r_m^l$ and the suboptimal power allocation $p_l^m$ of \eqref{opt:power_mno_alt}.

To build the datasets $\bm{D}$ for training and testing the model, we first generate the channel coefficients $\G^m$ according to the process described in subsection \ref{sec:eval1}, while for simplicity we use the fixed $\bar{\r}$ and $\bm{\sigma}^m$ for all data instances. In this subsection, we generate $2000$ problem instances of  \eqref{opt:power_mno_alt} under three different rate constraints (i.e., with rate function in the constraints as the CDMA approximation rate function, Shannon Capacity, and Bit Error Rate). These problem instances are used samples, with each having four users communicating using two subcarriers.
 For the labels to conduct supervised learning, we use the brute-force search (e.g., a naive grid search) or empirical knowledge to obtain the transmission rate transmission, and then use Algorithm \ref{alg:decentralizedAlg} to obtain power allocations to generate all the training datasets, testing datasets, and validation datasets. The testing and validation datasets are much smaller than the training datasets and are used for cross-validation and early stopping during the training phase. We randomly split the whole dataset in the proportion of $80\%$, $10\%$, and $10\%$  for training, testing, and validation, respectively.

 To configure the OpenRANet, we set the convolutional filter size to $3\times3$ and utilize the average pooling method. In the dense layers, we incorporate $3$ hidden layers, with each layer comprising $128$ neurons. The activation function employed is RELU. The learning rate is set at $1\times 10^{-3}$, and the weight decay is $1\times 10^{-4}$. We use the mean squared error between the labels (the ground truths of transmission rate and the optimal power allocation) and the outputs of the model as the loss function, as defined in (\ref{eq:loss}). To train the learning parameters, we employ the ADAM optimizer \cite{kingma2015adam} and perform the training for a maximum of $100$ epochs. To further improve the training performance and speed, instead of randomly selecting the initial learning parameters, we use a pre-training method to determine the initial learning parameters. To do this, we first use Algorithm \ref{alg:decentralizedAlg} to
generate a sufficiently large training dataset with local optimal solutions as labels and then use this training dataset to pre-train the learning parameters of the model.

\begin{figure}[htbp]
\centering
\vspace{-5mm}
\subfigure[]{
\begin{minipage}[t]{0.45\linewidth}
\centering
\includegraphics[width=1.6in]{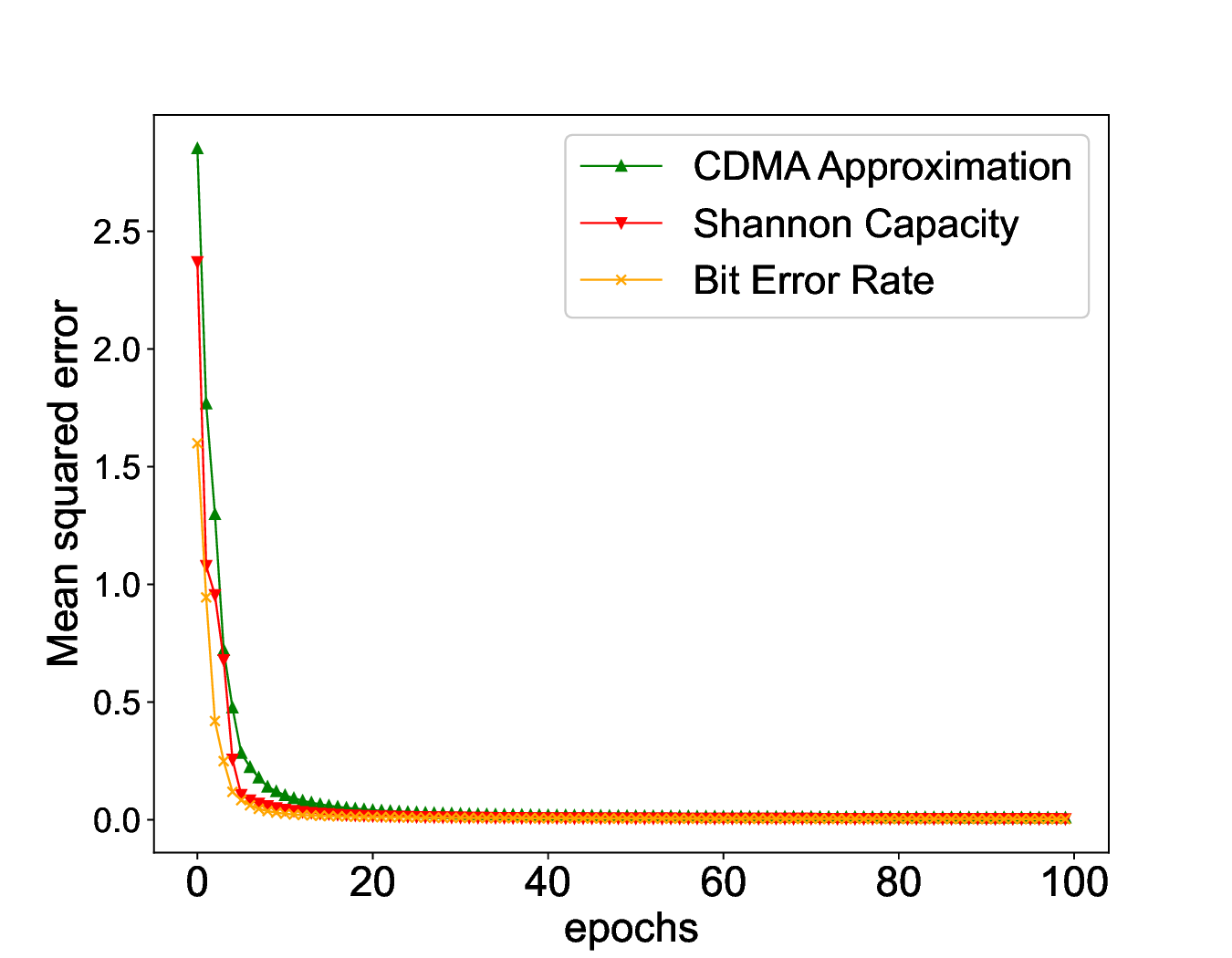}
\end{minipage}%
}%
\subfigure[]{
\begin{minipage}[t]{0.45\linewidth}
\centering
\includegraphics[width=1.6in]{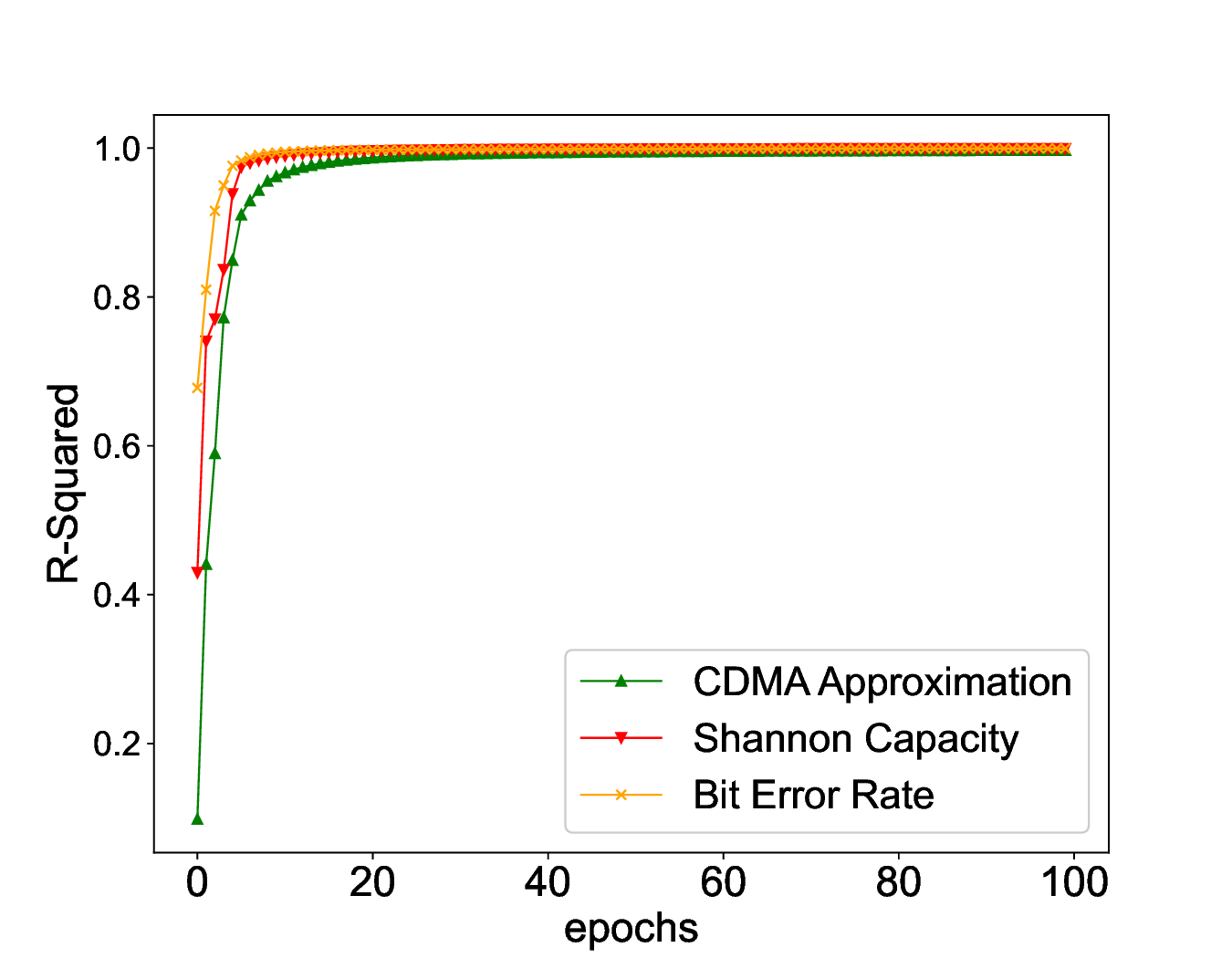}
\end{minipage}%
}%
\centering
\captionsetup{font={footnotesize}, justification=raggedright}
\caption{The mean squared error (a) and R-squared (b) of approximating optimal power allocation for the OpenRANet on the power minimization problem under different rate function constraints.}
\label{fig:loss}
\end{figure}

\begin{figure*}[htbp]
\centering
\vspace{-8mm}
\subfigure[CDMA Approximation-$r_1$]{
\begin{minipage}[t]{0.26\linewidth}
\centering
\includegraphics[width=1.6in]{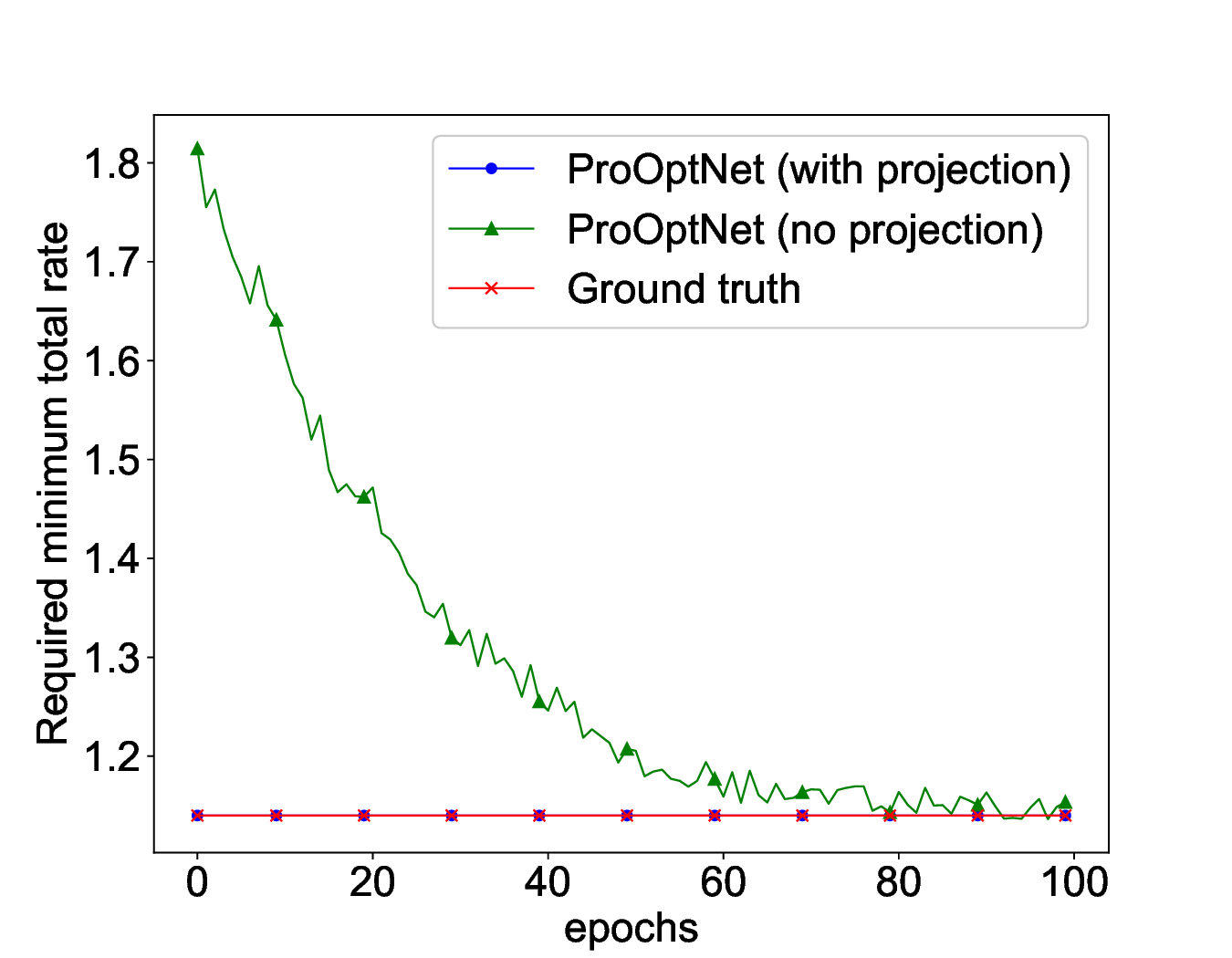}
\end{minipage}%
}%
\subfigure[Shannon Capacity-$r_1$]{
\begin{minipage}[t]{0.26\linewidth}
\centering
\includegraphics[width=1.6in]{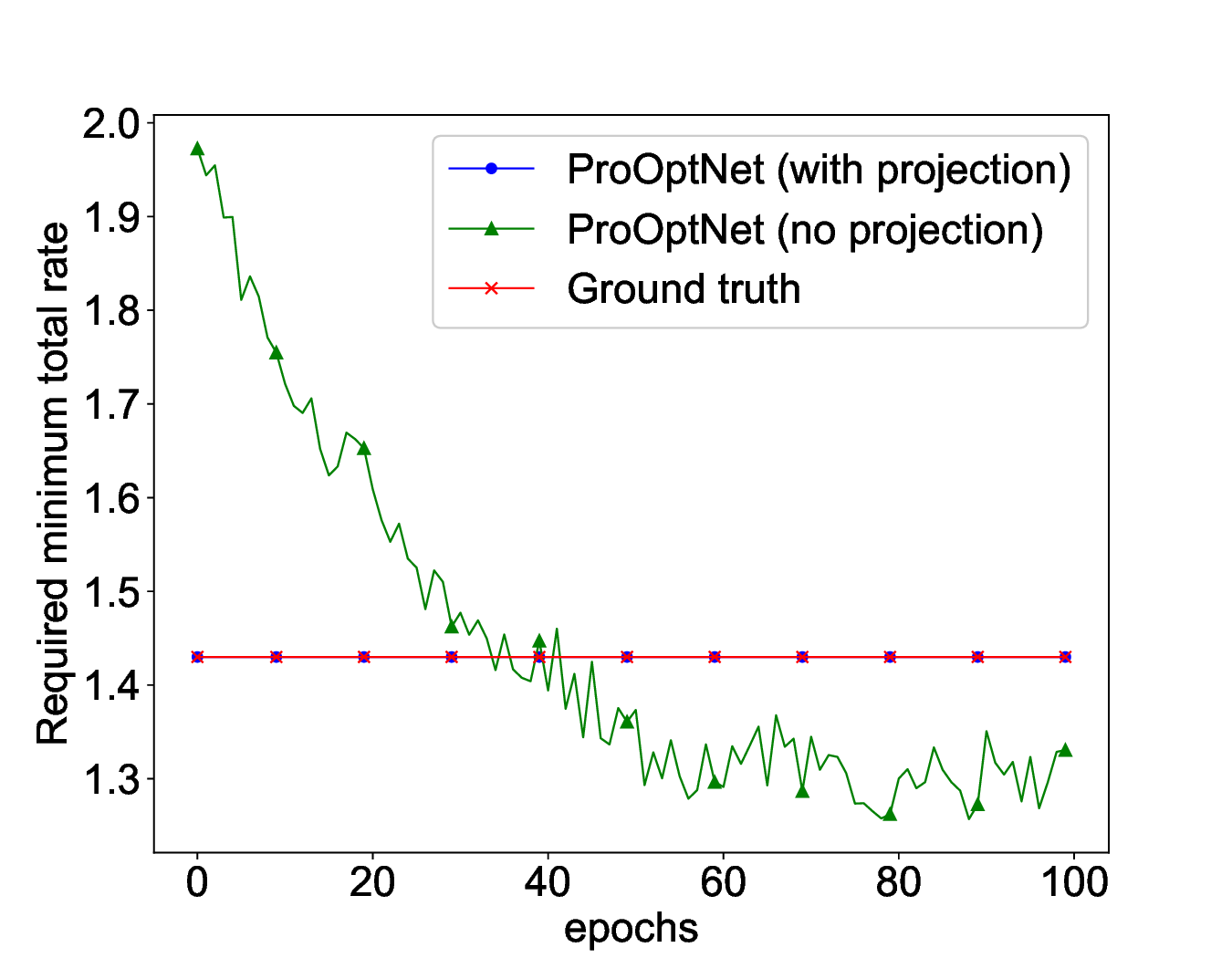}
\end{minipage}%
}%
\subfigure[Bit Error Rate-$r_1$]{
\begin{minipage}[t]{0.26\linewidth}
\centering
\includegraphics[width=1.6in]{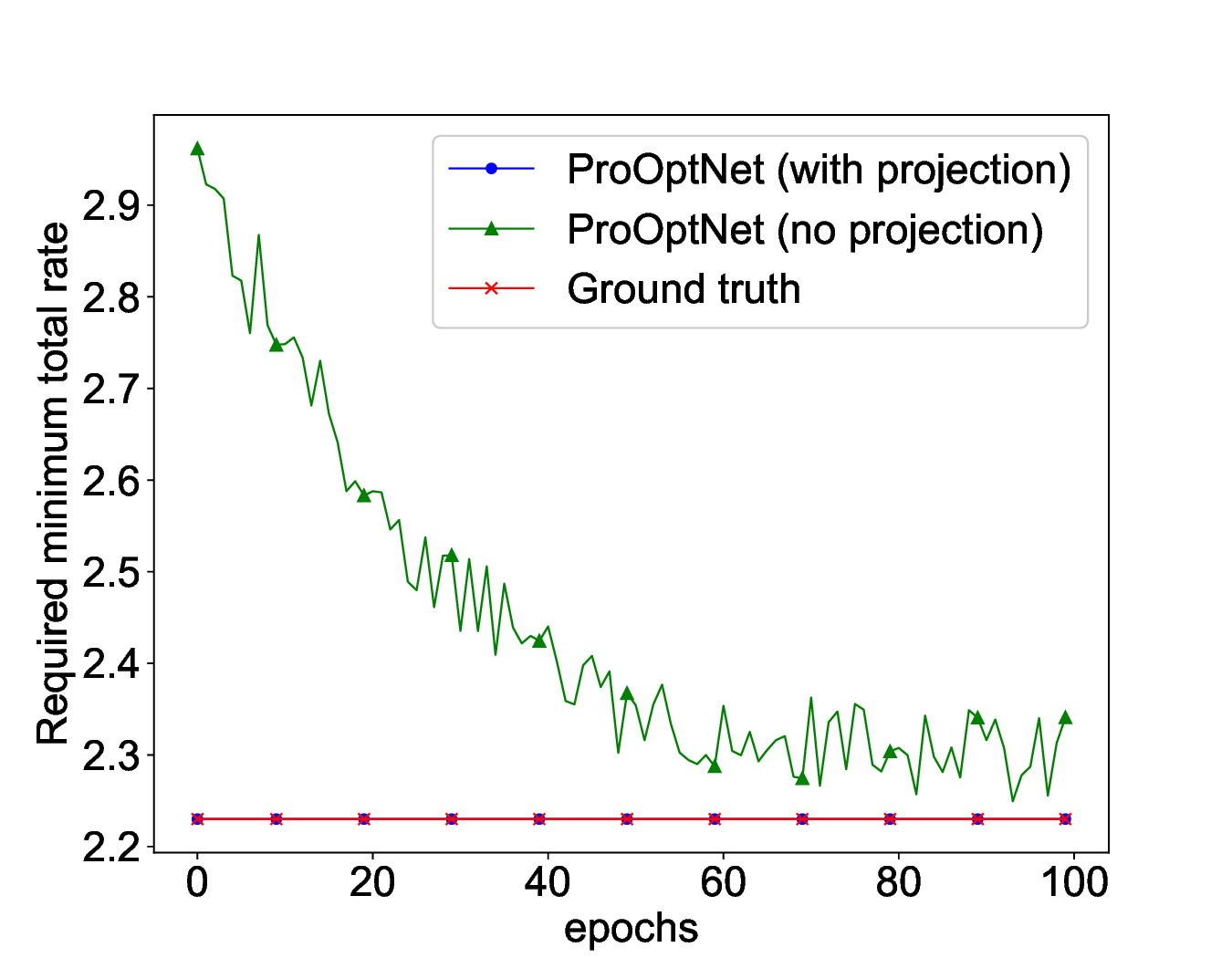}
\end{minipage}%
}%
\vspace{-3mm}

\subfigure[CDMA Approximation-$r_2$]{
\begin{minipage}[t]{0.26\linewidth}
\centering
\includegraphics[width=1.6in]{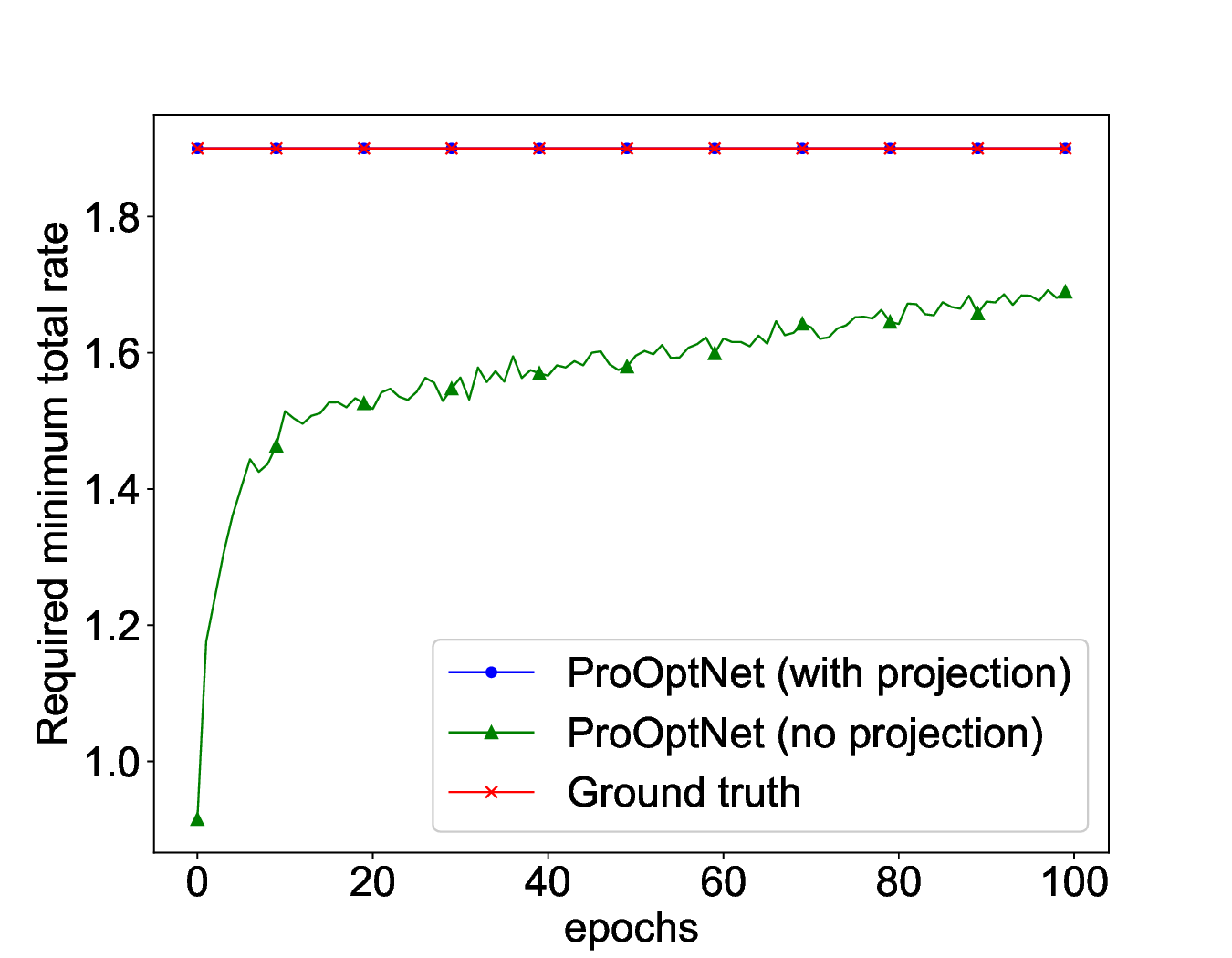}
\end{minipage}%
}%
\subfigure[Shannon Capacity-$r_2$]{
\begin{minipage}[t]{0.26\linewidth}
\centering
\includegraphics[width=1.6in]{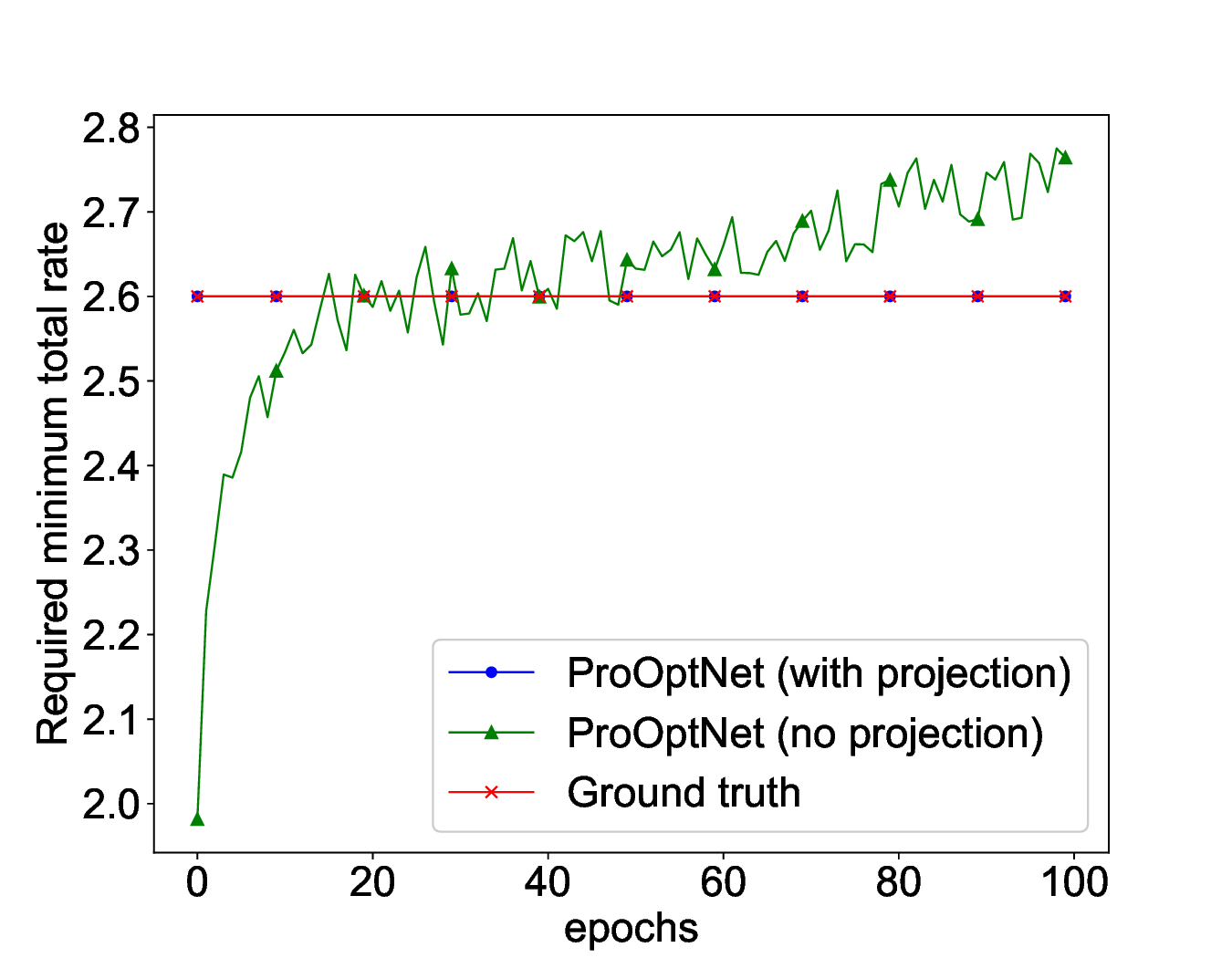}
\end{minipage}%
}%
\subfigure[Bit Error Rate-$r_2$]{
\begin{minipage}[t]{0.26\linewidth}
\centering
\includegraphics[width=1.6in]{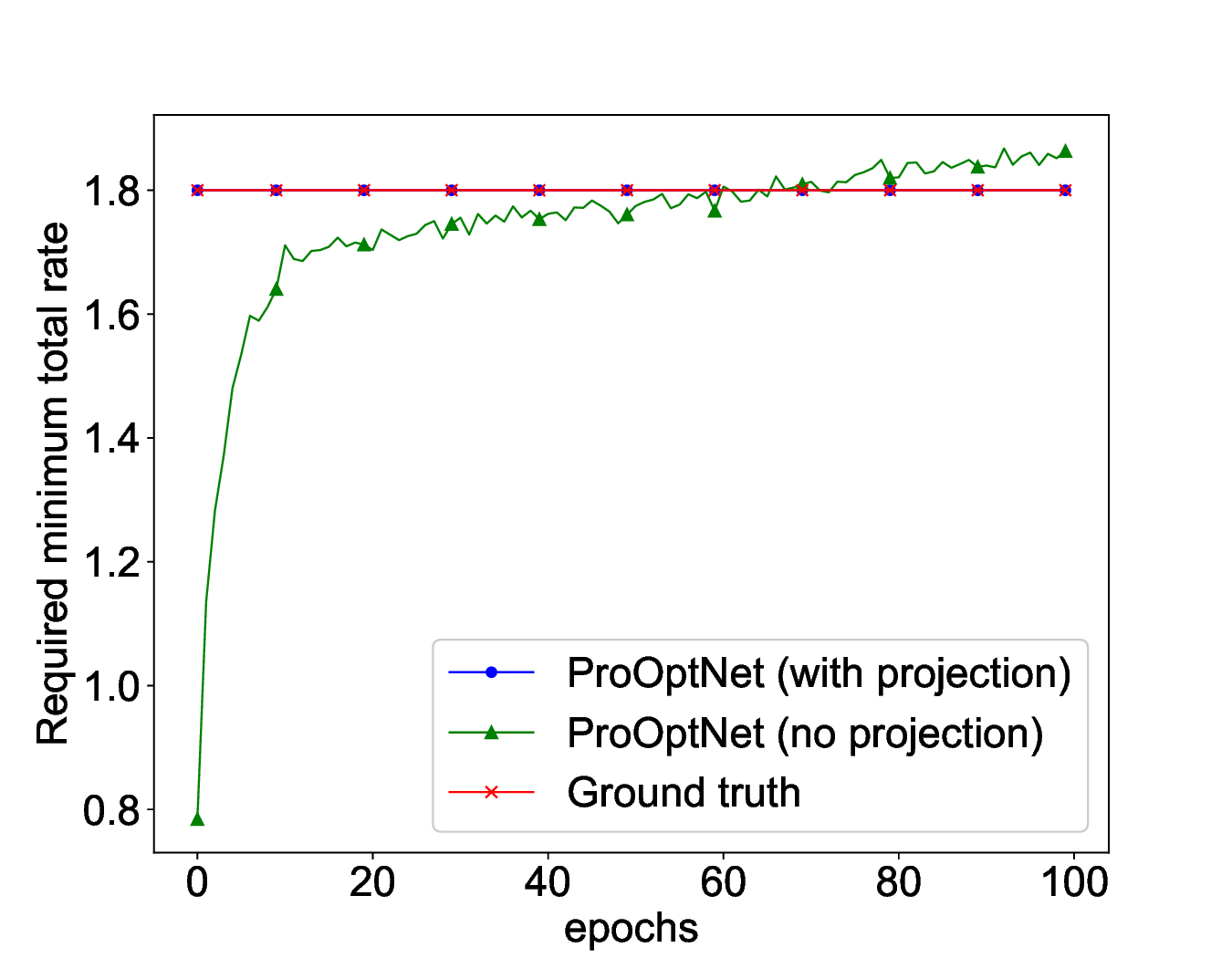}
\end{minipage}%
}%
\centering
\captionsetup{font={footnotesize}, justification=raggedright}
\caption{Comparison between the OpenRANet with and without the projection layer against the ground truth (the required minimum total transmission rate in the subcarriers) under different rate constraints.}
\label{fig:compare_projectL}
\end{figure*}

\begin{figure*}[htbp]
\centering
\vspace{-3mm}
\subfigure[CDMA Approximation]{
\begin{minipage}[t]{0.26\linewidth}
\centering
\includegraphics[width=1.6in]{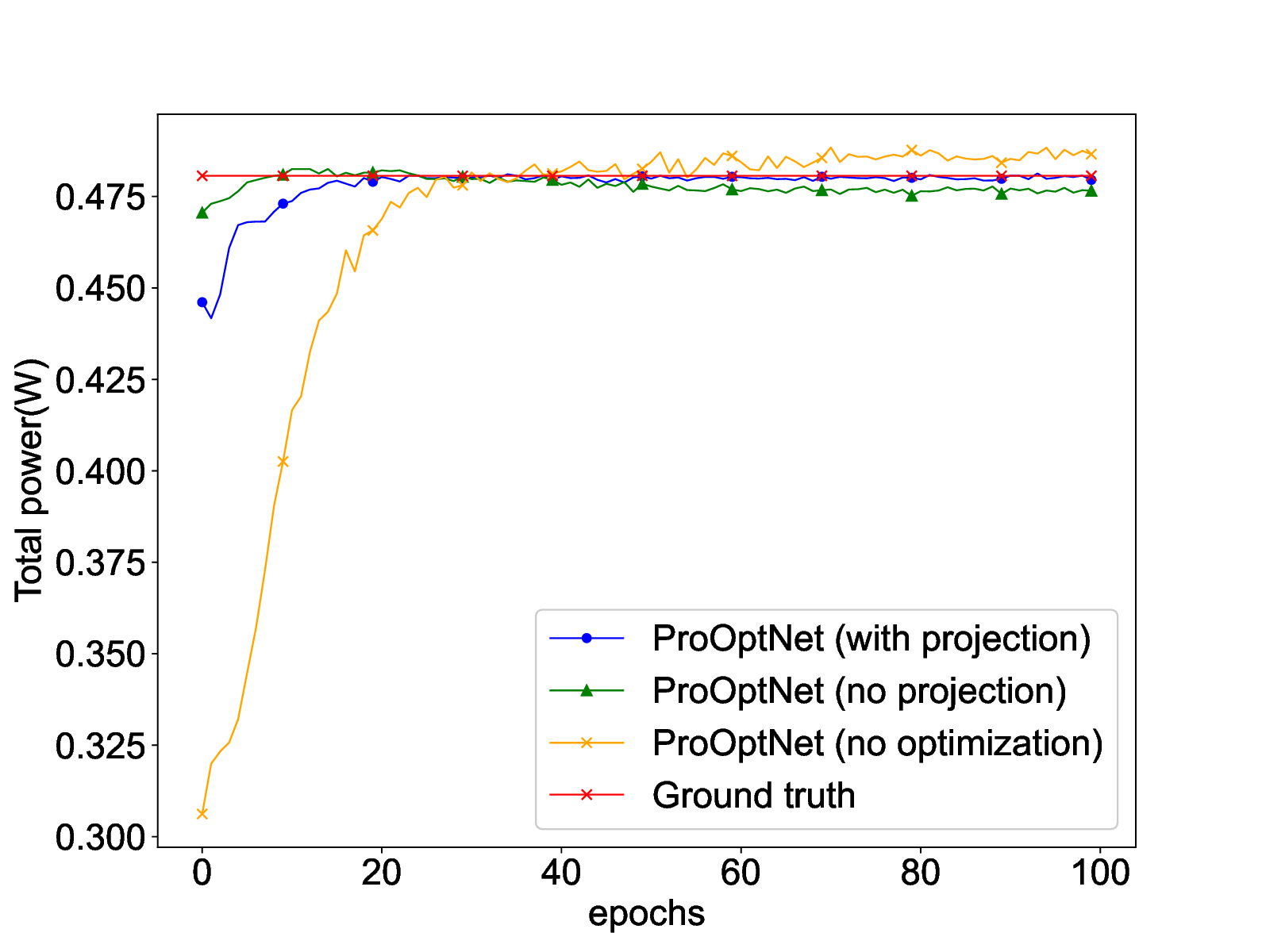}
\end{minipage}%
}%
\subfigure[Shannon Capacity]{
\begin{minipage}[t]{0.26\linewidth}
\centering
\includegraphics[width=1.6in]{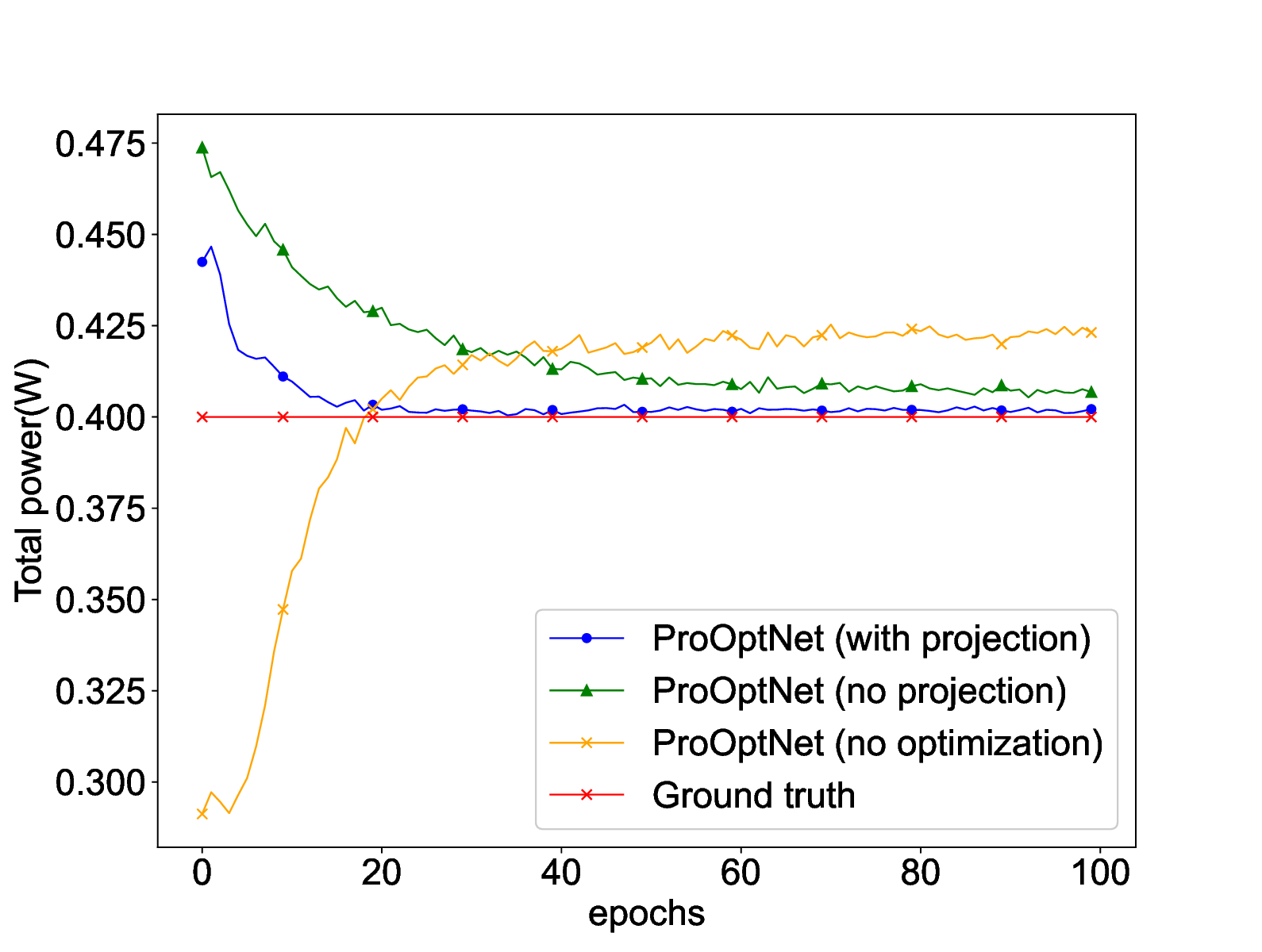}
\end{minipage}%
}%
\subfigure[Bit Error Rate]{
\begin{minipage}[t]{0.26\linewidth}
\centering
\includegraphics[width=1.6in]{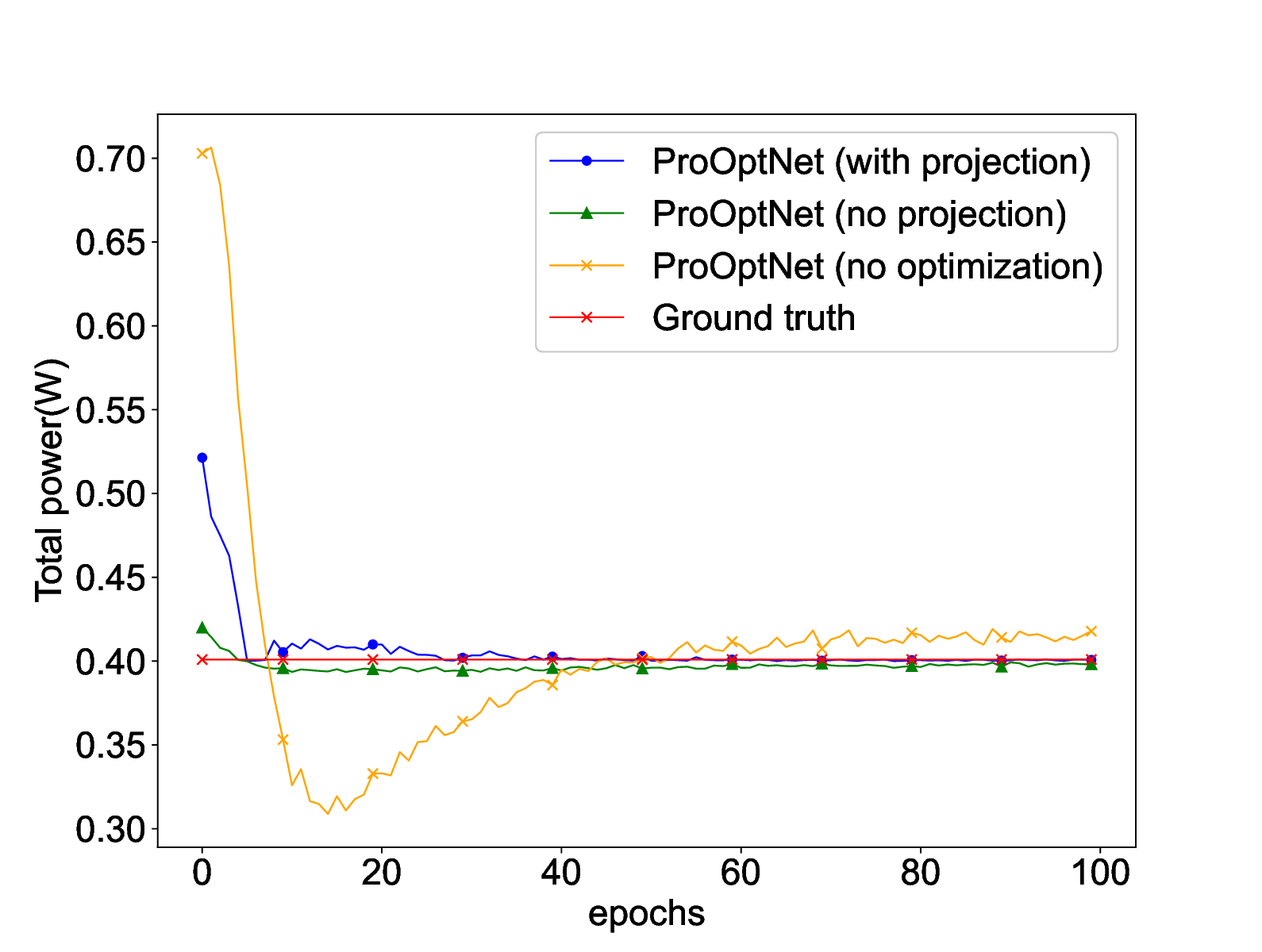}
\end{minipage}%
}%
\centering
\captionsetup{font={footnotesize}, justification=raggedright}
\caption{Comparison between the OpenRANet with and without the optimization layer against the ground truth (the minimum total power) under different rate constraints.}
\label{fig:compare_OptL}
\end{figure*}

\textcolor{black}{Fig. \ref{fig:loss} depicts the mean squared error \cite{chicco2021coefficient} (a) and the coefficient of determination (R-squared) \cite{chicco2021coefficient}  (b)  of approximating optimal power allocation over different epochs for the OpenRANet for power minimization under various rate function constraints.} This demonstrates that OpenRANet achieves high levels of accuracy and is an excellent fit for the data. To highlight the significance of the projection layer in the OpenRANet, we conduct a comparative analysis between the OpenRANet with and without the projection layer against the ground truth (the required minimum total transmission rate in the subcarriers). Note that the benefit of the projection layer is to ensure that the OpenRANet's output adheres to the constraints of the problem in \eqref{opt:power_mno_alt}. The results of this comparison under various rate constraints are illustrated in Fig. \ref{fig:compare_projectL}. We can see that the transmission rates of outputs of the OpenRANet with the projection layer always meet the transmission rate requirements. However, Fig. \ref{fig:compare_projectL} (b) and Fig. \ref{fig:compare_projectL} (d) indicate that the removal of the projection layer can potentially lead to a total transmission rate in the subcarrier that falls below the required minimum level.
Actually, inserting a projection layer in the middle of the model is better than projecting the output of the model, because the knowledge of the transmission rate requirements can be used to train the learning parameters as it can be difficult to derive a closed form of the power distribution for the transmission rate for projection.
In addition, we compare the OpenRANet with and without the optimization layer to the ground truth (the minimum total power) to showcase the significance of the optimization layer in the OpenRANet. The results of this comparison under different rate constraints are illustrated in Fig. \ref{fig:compare_OptL}. It is shown that the OpenRANet with the optimization layer outperforms the one without the optimization layer, as it achieves a smaller total power output for the problem in \eqref{opt:power_mno_alt}.

\vspace{-3mm}

\subsection{Performance Comparison}
In this subsection, we analyze problem samples that involve a higher number of users ($128$ users) and subcarriers ($16$ subcarriers). 
\textcolor{black}{We examine two categories of baselines:  optimization-based methods and machine-learning methods.
More details of the baselines are available in \href{https://arxiv.org/abs/2409.12964} {the ArXiv online version}.
The optimization-based baselines include the successive convex approximation (SCA) method, as discussed in \cite{pennanen2015decentralized}, and the branch and bound method derived from \cite{zhai2017rate} and \cite{balakrishnan1991branch}. The SCA method stands out for its efficiency, although it may not always lead to globally optimal solutions. Conversely, the Branch and Bound method is noted for its capability to ascertain globally optimal solutions, albeit with a potential compromise in efficiency. 
To design the SCA method for solving the proposed nonconvex power minimization problem, we emulate \cite{pennanen2015decentralized} by using the first-order Taylor expansion. We use $e^{\tilde{r}_l^m(k)} (\tilde{r}_l^m-\tilde{r}_l^m(k))-\sum_{m=1}^M e^{\tilde{r}_l^m(k)}$ to approximate the concave $-\sum_{m=1}^M e^{\tilde{r}_l^m}$ at $\tilde{r}_l^m(k)$, and then construct approximate convex subproblems. Iteratively solving these subproblems yields at least locally optimal solutions.
To design the  BnB method as baselines, we use Algorithm 1's solution for the upper bound and obtain the lower bound via relaxation of the closest concave envelope for the constraints' exponential functions. Leveraging the BnB mechanism in \cite{balakrishnan1991branch}, we can construct a BnB method to find the global solution for the proposed nonconvex problem.
For machine learning baselines, we refer to the deep neural network (DNN) model in \cite{camana2022deep}, graph neural networks (GNN) in \cite{eisen2020optimal}, and the deep belief network (DBN) in \cite{luo2019deep}. 
In designing the DNN model, we follow \cite{camana2022deep}, inputting problem parameters like channel gain, noise power, and the minimal transmission rate per user per subcarrier. The output, an approximate $r_l^m$, is used to solve convex subproblems, but we do not use these subproblems as layers for backpropagation to update the learning parameters.
For the GNN model, we consider the communication system as a graph $G(V, E)$, with nodes as users and edges as interference. Node features include weights, noise power $n_i$, and maximum power, while edge features represent channel gain $G_{i j}$. We then apply a GNN model to $G(V, E)$ and output the approximate solution, or optimal $p_l^m$.
The design of the DBN model is quite similar to the DNN model, but inputs only the system model's channel gain.}
 \begin{figure}
 \vspace{-5mm}
\centerline{\includegraphics[scale=0.28]{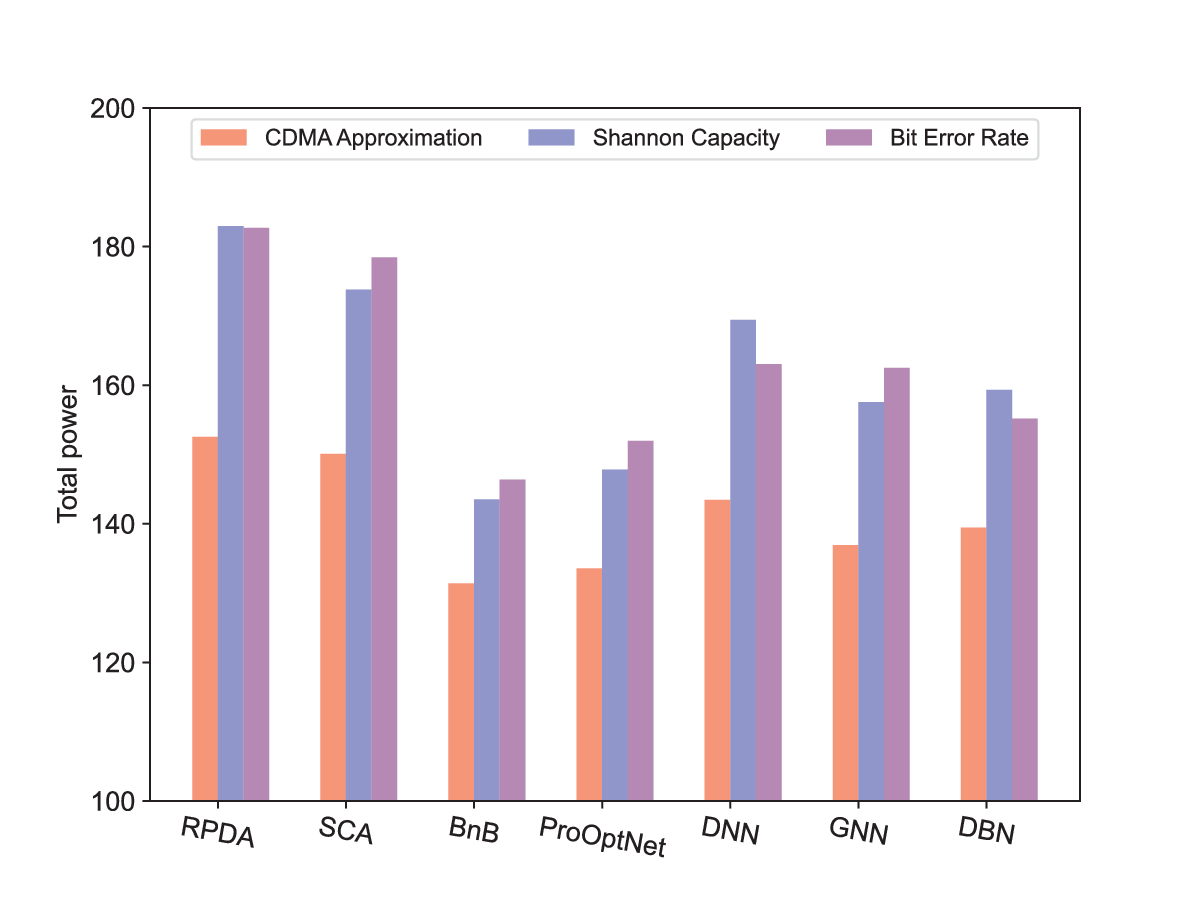}}
\captionsetup{font={footnotesize}, justification=raggedright}
\caption{The comparison of the total power for the proposed OpenRANet and other baselines.}
\label{fig:totalpower} 
\end{figure}
The experimental results are shown in Fig. \ref{fig:totalpower} and TABLE \ref{tab:tab1}. Fig. \ref{fig:totalpower} illustrates the comparison of the total power for the total power minimization problem under different rate constraints for the proposed OpenRANet and other baselines. The results of BnB serve as the ground truth, although its computational complexity is high. We can observe that the proposed OpenRANet is very close to the ground truth, while significantly outperforming the other baselines. 

In TABLE \ref{tab:tab1}, we compare the training time and the number of iterations required for the proposed model and the baselines. The deep learning baselines, such as DNN, GNN, and DBN, can obtain approximate solutions to the problem with only a few layers of iterations (configured as $8$ layers/iteration, as shown in TABLE \ref{tab:tab1}). However, they require significant time and resource costs for training the models in advance. On the other hand, optimization-based baselines, such as RPDA, SCA, and BnB, do not require pre-training the models at a cost but need a large number of iterations to obtain approximate solutions to the problem. We can observe that the training time for OpenRANet is significantly lower than that of other pure deep learning models and requires much fewer iterations than the optimization-based baselines to achieve optimal solutions. This demonstrates that integrating the intrinsic information of the optimization problem into a deep learning model can improve the accuracy of the model while reducing the difficulty and cost of training it.




\begin{table*}[]\footnotesize
\centering 
\renewcommand{\arraystretch}{1}
\begin{tabular}{c|cc|cc|cc}
\hline
\multirow{2}{*}{\textbf{Methods}} &
  \multicolumn{2}{c|}{\textbf{CDMA Approximation}} &
  \multicolumn{2}{c|}{\textbf{Shannon Capacity}} &
  \multicolumn{2}{c}{\textbf{Bit Error Rate}} \\ \cline{2-7} 
 &
  \multicolumn{1}{l|}{\textbf{Training time(s)}} &
  \multicolumn{1}{l|}{\textbf{Iterations}} &
  \multicolumn{1}{l|}{\textbf{Training time(s)}} &
  \multicolumn{1}{l|}{\textbf{Iterations}} &
  \multicolumn{1}{l|}{\textbf{Training time(s)}} &
  \multicolumn{1}{l}{\textbf{Iterations}} \\ \hline 
\textbf{RPDA}       & \multicolumn{1}{c|}{--}    & 7.8*10e3 & \multicolumn{1}{c|}{--}    & 8.4*10e3 & \multicolumn{1}{c|}{--}    & 7.6*10e3 \\
\textbf{SCA}       & \multicolumn{1}{c|}{--}    & 1.1*10e3 & \multicolumn{1}{c|}{--}    & 1.4*10e3 & \multicolumn{1}{c|}{--}    & 1.3*10e3 \\ [0.5ex]
\textbf{BnB}       & \multicolumn{1}{c|}{--}    & 5.3*10e3 & \multicolumn{1}{c|}{--}    & 6.3*10e3 & \multicolumn{1}{c|}{--}    & 5.7*10e3 \\ [0.5ex]
\textbf{OpenRANet} & \multicolumn{1}{c|}{68.34} & 1.3*10e2 & \multicolumn{1}{c|}{78.43} & 1.4*10e2 & \multicolumn{1}{c|}{72.87} & 0.8*10e2 \\ [0.5ex]
\textbf{DNN}       & \multicolumn{1}{c|}{88.45} & 8        & \multicolumn{1}{c|}{93.67} & 8        & \multicolumn{1}{c|}{90.21} & 8        \\ [0.5ex]
\textbf{GNN}       & \multicolumn{1}{c|}{79.86} & 8        & \multicolumn{1}{c|}{80.23} & 8        & \multicolumn{1}{c|}{81.09} & 8        \\ [0.5ex]
\textbf{DBN}       & \multicolumn{1}{c|}{86.23} & 8        & \multicolumn{1}{c|}{88.99} & 8        & \multicolumn{1}{c|}{80.23} & 8        \\ [0.5ex] \hline
\end{tabular}
\captionsetup{font={footnotesize}, justification=raggedright}
\caption{The training time and number of iterations for the proposed OpenRANet and other baselines.}
\label{tab:tab1}
\end{table*}

\section{Conclusion}\label{sec:conclusion}
This paper addresses the issue of minimizing total power consumption while meeting transmission rate requirements in the open RAN system. This problem is challenging due to its nonconvex nature, making it difficult to achieve optimality. Additionally, the existence of coupling rate constraints complicates the solution distribution, leading to increased energy consumption in the networks. To tackle these obstacles, we initially introduce a primal-dual algorithm that utilizes the unique log-convexity property of standard interference functions to coordinate power allocation among cooperating subcarriers. Furthermore, we propose an optimization-based deep learning model called OpenRANet, which integrates constraint information and convex subproblems into the deep learning model for subcarrier and power allocation for minimizing total power consumption. Our numerical experiments indicate that OpenRANet outperforms many cutting-edge strategies. 
The OpenRANet framework can serve as a foundation, and we plan to extend it in the future to accommodate multi-cell systems and additional system power consumption requirements, such as the energy required for signal processing, cooling, and neural network processing. Another consideration is adapting the model to scenarios where the traffic patterns change over time. To address this issue, we can integrate strategies such as transfer learning and incremental learning into OpenRANet. We believe this issue deserves further investigation in future studies.

\appendix

\subsection{Proof of Lemma~\ref{p_standard}}
By the positivity of the transmit power $\mathbf{p}^m$ and the achievable rate $r_l^m$, the positivity of $I_l^m(\mathbf{p}^m)$ holds trivially.

Let us first prove the monotonicity of $I_l^m(\mathbf{p}^m)$. To avoid undue clutter here, we use $f_l^m(\scalemath{0.9}{\mathsf{SINR}}_l^m)$ instead of $f_l^m(\scalemath{0.9}{\mathsf{SINR}}_l^m( \mathbf{p}^m))$.

For $j=l$, we have
\begin{align*}
\frac{\partial I_l^m(\mathbf{p}^m)}{\partial p_l^m}
=\frac{ f_l^m(\scalemath{0.9}{\mathsf{SINR}}_l^m)-\frac{\displaystyle\partial f_l^m(\scalemath{0.9}{\mathsf{SINR}}_l^m)}{\displaystyle\partial \scalemath{0.9}{\mathsf{SINR}}_l^m}\scalemath{0.9}{\mathsf{SINR}}_l^m}{ f_l^m(\scalemath{0.9}{\mathsf{SINR}}_l^m)^2}r_l^m
\geq 0,
\end{align*}

since $\frac{\partial f_l^m(\scalemath{0.9}{\mathsf{SINR}}_l^m)}{\partial\scalemath{0.9}{\mathsf{SINR}}_l^m}\leq \frac{f_l^m(\scalemath{0.9}{\mathsf{SINR}}_l^m)}{\scalemath{0.9}{\mathsf{SINR}}_l^m}$ holds by assumption.

For $j\neq l$, we have
\begin{align*}
    &\frac{\partial I_l^m(\mathbf{p}^m)}{\partial p_j^m} =
    -\frac{\frac{\displaystyle\partial f_l^m(\scalemath{0.9}{\mathsf{SINR}}_l^m)}{\displaystyle\partial \scalemath{0.9}{\mathsf{SINR}}_l^m}\frac{\displaystyle\partial \scalemath{0.9}{\mathsf{SINR}}_l^m}{\displaystyle\partial p_j^m} }{ f_l^m(\scalemath{0.9}{\mathsf{SINR}}_l^m)^2} r_l^mp_l^m\nonumber\\
     =&\frac{\frac{\displaystyle\partial f_l^m(\scalemath{0.9}{\mathsf{SINR}}_l^m)}{\displaystyle\partial \scalemath{0.9}{\mathsf{SINR}}_l^m}G_{lj}^mG_{ll}^m(p_l^m)^2 }{ f_l^m(\scalemath{0.9}{\mathsf{SINR}}_l^m)^2( \sum_{n\neq l}G_{ln}^mp_n^m+\sigma_l^m)^2} {r_l^m}\geq 0.
\end{align*}
Therefore, $\mathbf{I}^m(\mathbf{p}^m)$ is a monotonically increasing function in $\mathbf{p}^m$ for all $m$.

Next, let us prove the scalability of $I_l^m(\mathbf{p}^m)$. For any $\alpha>1$, we have
\begin{align*}
\scalemath{0.9}{\mathsf{SINR}}_l^m(\alpha \mathbf{p}^m)&\!=\!\frac{\alpha G_{ll}^mp_l^m}{\alpha \sum_{j \neq l}G_{lj}^mp_j^m+\sigma_l^m}\!=\!\frac{ G_{ll}^mp_l^m}{ \sum_{j \neq l}G_{lj}^mp_j^m\!+\!\sigma_l^m/\alpha}
\nonumber\\
 &> \frac{ G_{ll}^mp_l^m}{ \sum_{j \neq l}G_{lj}^mp_j^m+\sigma_l^m}=\scalemath{0.9}{\mathsf{SINR}}_l^m( \mathbf{p}^m).
\end{align*}
Since $\frac{\partial f_l^m(\scalemath{0.9}{\mathsf{SINR}}_l^m)}{\partial\scalemath{0.9}{\mathsf{SINR}}_l^m}>0$ holds by assumption, which implies $f_l^m(\scalemath{0.9}{\mathsf{SINR}}_l^m(\alpha \mathbf{p}^m))> f_l^m(\scalemath{0.9}{\mathsf{SINR}}_l^m(\mathbf{p}^m))$, then we have
\begin{align*}
I_l^m(\alpha\mathbf{p}^m) &= \frac{ {r_l^m}} {f_l^m(\alpha\scalemath{0.9}{\mathsf{SINR}}_l^m(\mathbf{p}^m))}\alpha p_l^m\nonumber\\
&< \alpha\frac{ r_l^m }{f_l^m(\scalemath{0.9}{\mathsf{SINR}}_l^m(\mathbf{p}^m))}p_l^m=\alpha I_l^m(\mathbf{p}^m).
\end{align*}

Therefore, $I_l^m(\mathbf{p}^m)$ is a standard interference function.

\subsection{Proof of Theorem~\ref{theorem2}}
Suppose that the rate constraint of a specific user $l$ is not tight at optimality, i.e.,
\begin{equation}\label{eq:rate_sum_ineq}
\displaystyle \sum_{m=1}^M {r_l^m}^\star = \sum_{m=1}^M f_l^m(\scalemath{0.9}{\mathsf{SINR}}_l^m({\mathbf{p}^m}^\star))> \bar{r}_l.
\end{equation}
Since $\frac{\partial f_l^m(\scalemath{0.9}{\mathsf{SINR}}_l^m(\mathbf{p}^m))}{\partial\scalemath{0.9}{\mathsf{SINR}}_l^m(\mathbf{p}^m)}>0$ holds, we can verify that the left-hand side of inequality~\eqref{eq:rate_sum_ineq} is strictly increasing in $p_l^m$ and strictly decreasing in $p_j^m$ for all $j \neq l$. If we reduce $p_l^m$ by a sufficiently small amount $\epsilon>0$ without violating~\eqref{eq:rate_sum_ineq}, it turns out that $\scalemath{0.9}{\mathsf{SINR}}_j^m(\mathbf{p}^m)$ for all $j \neq l$ will increase and still satisfy the rate constraints. This contradicts the assumption that the powers in (\ref{eq:rate_sum_ineq}) are optimal.

\subsection{Proof of Lemma~\ref{thm3}}
For the transmission rate function $f_l^m(\scalemath{0.9}{\mathsf{SINR}}_l^m(\mathbf{p}^m))$ of the $l$-th user on the $m$-th subcarrier, 
let $\tilde{\mathbf{p}}^m=\log \mathbf{p}^m$ and define $g_l^m(\tilde{\mathbf{p}}^m)$ by $g_l^m(\tilde{\mathbf{p}}^m)=\log f_l^m(\scalemath{0.9}{\mathsf{SINR}}_l^m(e^{\tilde{\mathbf{p}}^m)})$.
By the chain rule of differentiation, the derivative of $g_l^m$ with respect to $\tilde{\mathbf{p}}^m$ is given by
\begin{align*}
\frac{\partial g_l^m(\tilde{\mathbf{p}}^m)}{\partial \tilde{p}^m_j}
= &  \displaystyle\frac{\partial f_l^m(\scalemath{0.9}{\mathsf{SINR}}_l^m(e^{\tilde{\mathbf{p}}^m}))}{\partial \scalemath{0.9}{\mathsf{SINR}}_l^m(e^{\tilde{\mathbf{p}}^m})} \cdot \frac{\partial \scalemath{0.9}{\mathsf{SINR}}_l^m(e^{\tilde{\mathbf{p}}^m})}{\partial \tilde{p}_j^m }\nonumber\\
&\cdot\frac{1}{f_l^m(\scalemath{0.9}{\mathsf{SINR}}_l^m(e^{\tilde{\mathbf{p}}^m}))}.
\end{align*}
Let us use $s$ to substitute $\scalemath{0.9}{\mathsf{SINR}}_l^m(e^{\tilde{\mathbf{p}}^m})$ for brevity in the following part.
Then the entries of the Hessian matrix $\nabla^2 g_l^m(\tilde{\mathbf{p}}^m)$ are derived as follows.

For $j=l$,
\begin{align*}
  &(\nabla^2 g_l^m(\tilde{\mathbf{p}}^m))_{ll}=\frac{\partial^2 g_l^m(\tilde{\mathbf{p}}^m)}{\partial (\tilde{p}_l^m)^2} \nonumber\\
   = & \left(\frac{\partial^2 f_l^m(s)}{\partial s^2}f_l^m(s)-(\frac{\partial f_l^m(s)}{\partial s})^2\right)\left(\frac{\partial s}{\partial \tilde{p}_l^m}\right)^2\frac{1}{f_l^m(s)^2}  \nonumber
\\
    &+\frac{\partial f_l^m(s)}{\partial s}\frac{\partial^2 s }{\partial (\tilde{p}_l^m)^2}\frac{1}{f_l^m(s)}.
\end{align*}

For $j\neq l$,
\begin{align*}
&(\nabla^2 g_l^m(\tilde{\mathbf{p}}^m))_{lj}=\frac{\partial^2 g_l^m(\tilde{\mathbf{p}}^m)}{\partial (\tilde{p}_j^m)^2}\nonumber
 \\
=&\left(\frac{\partial^2 f_l^m(s)}{\partial s^2}f_l^m(s)-(\frac{\partial f_l^m(s)}{\partial s})^2\right)\frac{\partial s}{\partial \tilde{p}_j^m}\frac{\partial s}{\partial \tilde{p}_l^m}\frac{1}{f_l^m(s)^2} \nonumber
 \\
  &+\frac{\partial f_l^m(s)}{\partial s}\frac{\partial^2 s }{\partial \tilde{p}_j^m \partial \tilde{p}_l^m}\frac{1}{f_l^m(s)}.
\end{align*}
Now let us show that if \eqref{eq:thm3_1} holds, then the Hessian matrix $\nabla^2 g_l^m(\tilde{\mathbf{p}}^m)$ is indeed negative definite.

For any real vector $\mathbf{z}$, we have
\begin{align}\label{eq:concavevector2}
&\displaystyle \mathbf{z}^\top\nabla^2 g_l^m(\tilde{\mathbf{p}}^m)\mathbf{z}
= \displaystyle \frac{\frac{\partial^2 f_l^m}{\partial y^2}f_l^my^2-\left(\frac{\partial f_l^m}{\partial y}\right)^2y^2+\frac{\partial f_l^m}{\partial y}f_l^my}{(f_l^m)^2}
\displaystyle \nonumber\\
&\cdot\left(z_l-\sum_{j\neq l}\frac{G_{li}^mp_j^m}{\sum_{i\neq l}G_{li}^mp_i^m+\sigma_l^m}z_j\right)^2\nonumber\\
&\displaystyle + \frac{\frac{\partial f_l^m}{\partial y}f_l^my}{(f_l^m)^2}\Bigg[ \Bigg(\sum_{j\neq l}\frac{G_{lj}^mp_j^m}{\sum_{i\neq l}G_{li}^mp_i^m+\sigma_l^m}z_j\Bigg)^2
 \nonumber\\
&-\sum_{j\neq l}\frac{G_{lj}^mp_j^m}{\sum_{i\neq l}G_{li}^mp_i^m+\sigma_l^m}z_j^2\Bigg]\leq0.
\end{align} 
It is observed that the first term in~\eqref{eq:concavevector2} is negative as long as \eqref{eq:thm3_1} holds.
The second term is negative because $\frac{\partial f_l^m}{\partial y}f_l^my\geq 0$ and the other parts correspond to the negative definite property of the Hessian matrix of the concave function $\log \scalemath{0.9}{\mathsf{SINR}}_l^m(e^{\tilde{\mathbf{p}}^m})$.

\subsection{Proof of Theorem~\ref{th:dual}}\label{theo2}
Let us introduce the nonnegative Lagrangian multipliers $\lambda_l^m$ corresponding to the rate constraints, and express the Lagrangian function as:
\begin{align}
L(\mathbf{\tilde{p}}^m, \boldsymbol{\lambda}^m)= \displaystyle \sum_{l=1}^L\sum_{m=1}^M e^{\tilde{p}_l^m}\! -\! 
\sum_{l=1}^L\sum_{m=1}^M \lambda_l^m \log\frac{\hat{r}_l^m}{f_l^m(\scalemath{0.9}{\mathsf{SINR}}_l^m(e^{\tilde{\mathbf{p}}^m}))}.
\end{align}
Using stationarity of the Lagrangian for all $l$ and $m$, we have
\begin{align}\label{eq:th4p}
  &\frac{\partial L(\mathbf{\tilde{p}}^m, \boldsymbol{\lambda}^m)}{\partial \tilde{p}_l^m}
  \!=\!\!\ e^{\tilde{p}_l^m}\!\!\!-\!\frac{\scalemath{0.9}{\mathsf{SINR}}_l^m(e^{\tilde{\mathbf{p}}^m}) }{f_l^m(\scalemath{0.9}{\mathsf{SINR}}_l^m(e^{\tilde{\mathbf{p}}^m}))}\frac{\partial f_l^m(\scalemath{0.9}{\mathsf{SINR}}_l^m(e^{\tilde{\mathbf{p}}^m}))}{\partial \scalemath{0.9}{\mathsf{SINR}}_l^m(e^{\tilde{\mathbf{p}}^m})}\lambda_l^m  \nonumber
\\
& +\displaystyle\sum_{j\neq l} \frac{\scalemath{0.9}{\mathsf{SINR}}_j^{m^2}(e^{\tilde{\mathbf{p}}^m})\frac{G_{jl}^me^{\tilde{p}_l^m}}{G_{ll}^me^{\tilde{p}_j^m}}}{f_j^m(\scalemath{0.9}{\mathsf{SINR}}_j^m(e^{\tilde{\mathbf{p}}^m}))}\frac{\partial f_j^m(\scalemath{0.9}{\mathsf{SINR}}_j^m(e^{\tilde{\mathbf{p}}^m}))}{\partial \scalemath{0.9}{\mathsf{SINR}}_j^m(e^{\tilde{\mathbf{p}}^m})}\lambda_j^m
 =\ 0.
\end{align}
Changing the variables from $\tilde{p}_l^m$ and $\hat{r}_l^m$ back to $p_l^m$ and $r_l^m$ respectively, we obtain~\eqref{eq:thm4_1} - \eqref{eq:thm4_3}.

\subsection{Proof of Corollary 1}
Suppose for all $l$ and $m$ that the transmission rate function is defined as $f_l^m(\scalemath{0.9}{\mathsf{SINR}}_l^m(\mathbf{p}^m)) =\mathsf{SINR}_l^m(\mathbf{p}^m)$, it is obvious that $I_l^m(\mathbf{p}^m)$ is a standard interference function.

    Suppose the transmission rate function is defined as $f_l^m(\scalemath{0.9}{\mathsf{SINR}}_l^m(\mathbf{p}^m)) = \log(1+\scalemath{0.9}{\mathsf{SINR}}_l^m(\mathbf{p}^m))$.
It is easy to verify that
\begin{align*}
    \frac{\partial f_l^m(\scalemath{0.9}{\mathsf{SINR}}_l^m(\mathbf{p}^m))}{\partial\scalemath{0.9}{\mathsf{SINR}}_l^m(\mathbf{p}^m)} = \frac{1}{1+\scalemath{0.9}{\mathsf{SINR}}_l^m(\mathbf{p}^m)} \geq 0,
\end{align*}
and that
\begin{align*}
    \frac{\partial f_l^m(\scalemath{0.9}{\mathsf{SINR}}_l^m(\mathbf{p}^m))}{\partial\scalemath{0.9}{\mathsf{SINR}}_l^m(\mathbf{p}^m)}& = \frac{1}{1+\scalemath{0.9}{\mathsf{SINR}}_l^m(\mathbf{p}^m)}\nonumber\\
    &\leq  \frac{\log(1+\scalemath{0.9}{\mathsf{SINR}}_l^m(\mathbf{p}^m))}{\scalemath{0.9}{\mathsf{SINR}}_l^m(\mathbf{p}^m))}\nonumber\\
   & =\frac{f_l^m(\scalemath{0.9}{\mathsf{SINR}}_l^m(\mathbf{p}^m))}{\scalemath{0.9}{\mathsf{SINR}}_l^m(\mathbf{p}^m)}.
\end{align*}
The last step follows since $x/(1+x)\leq\log(1+x)$ for $x>0$.

As shown, since $f_l^m(\scalemath{0.9}{\mathsf{SINR}}_l^m(\mathbf{p}^m))$ satisfies assumptions in Lemma 2, we can conclude that by definition, $I_l^m(\mathbf{p}^m)=\frac{{r_l^m}}{\log(1+\scalemath{0.9}{\mathsf{SINR}}_l^m(\mathbf{p}^m))}p_l^m$ is a standard interference function.

Suppose the data rate function is given by $f_l^m(\scalemath{0.9}{\mathsf{SINR}}_l^m(\mathbf{p}^m)) =R(1-2Q\sqrt{\mathsf{SINR}_l^m(\mathbf{p}^m)})$, we can verify assumptions in Lemma 2 by:
\begin{equation}
\displaystyle\frac{Re^{-\mathsf{SINR}_l^m(\mathbf{p}^m)/2}}{\sqrt{2\pi \mathsf{SINR}_l^m(\mathbf{p}^m)}} \leq \frac{R(1-2Q(\sqrt{\mathsf{SINR}_l^m(\mathbf{p}^m)})}{\mathsf{SINR}_l^m(\mathbf{p}^m)},
\end{equation}
which is always equivalent to:
\begin{equation}\label{ieq:der1}
\displaystyle Q(\sqrt{\mathsf{SINR}_l^m(\mathbf{p}^m)}) \leq \frac{1}{2}-\frac{\mathsf{SINR}_l^m(\mathbf{p}^m)}{2\sqrt{2\pi \mathsf{SINR}_l^m(\mathbf{p}^m)}}e^{-\mathsf{SINR}_l^m(\mathbf{p}^m)/2}.
\end{equation}

\subsection{Proof of Corollary 2}
Suppose for all $l$ and $m$ that the transmission rate function is defined as $f_l^m(\scalemath{0.9}{\mathsf{SINR}}_l^m(\mathbf{p}^m)) =\mathsf{SINR}_l^m(\mathbf{p}^m)$, it is obvious that $f_l^m(\scalemath{0.9}{\mathsf{SINR}}_l^m(\mathbf{p}^m)) $ is log-concave.

Suppose the data rate function is $f_l^m(\mathsf{SINR}_l^m(\mathbf{p}^m)) = \log \left(1+\mathsf{SINR}_l^m(\mathbf{p}^m)\right)$, we can verify equation (6) in Lemma 3 to be true by:
\begin{equation}
\begin{array}{ll}
&\displaystyle -\frac{s\log(1+s)}{(1+s)^2}-\frac{\log(s)}{(1+s)^2}+\frac{\log(1+s)}{1+s}\leq0,
\end{array}
\end{equation}
for any positive power allocations, i.e., $s>0$.
Suppose the data rate function is $f_l^m(\mathsf{SINR}_l^m(\mathbf{p}^m))=R\left(1-2Q\left(\sqrt{\mathsf{SINR}_l^m(\mathbf{p}^m)}\right)\right)$, we can verify  equation (6) in Lemma 3 by:
\begin{equation}
\begin{array}{ll}
 \displaystyle-\frac{s+1}{2s\sqrt{2\pi s}}R^2e^{-s/2}(1-2Q(\sqrt{s})s-\left(\frac{R}{\sqrt{2\pi s}}e^{-s/2}\right)^2s\\
  \quad \quad \displaystyle+\frac{R}{\sqrt{2\pi s}}e^{-s/2}R(1-2Q(\sqrt{s})\leq0,
\end{array}
\end{equation}
which is equivalent to:
\begin{equation}\label{ieq:der2}
\begin{array}{ll}
&\displaystyle \frac{1-\mathsf{SINR}_l^m(e^{\tilde{\mathbf{p}}^m})}{2}\left(1-2Q\left(\sqrt{\mathsf{SINR}_l^m(e^{\tilde{\mathbf{p}}^m})}\right)\right) \\
\leq & \displaystyle \frac{\mathsf{SINR}_l^m(e^{\tilde{\mathbf{p}}^m})}{\sqrt{2\pi \mathsf{SINR}_l^m(e^{\tilde{\mathbf{p}}^m})}}e^{-\mathsf{SINR}_l^m(e^{\tilde{\mathbf{p}}^m})/2}.
\end{array}
\end{equation}
Note that~\eqref{ieq:der2} is satisfied when $\mathsf{SINR}_l^m(e^{\tilde{\mathbf{p}}^m}) \geq 1$.
On the other hand, if $0< \mathsf{SINR}_l^m(\mathbf{p}^m) <1$, then~\eqref{ieq:der2} is equivalent to:
\begin{equation}\label{ieq:der2f}
\begin{array}{ll}
&\displaystyle  Q\left(\sqrt{\mathsf{SINR}_l^m(\mathbf{p}^m)}\right) \\
\geq &\displaystyle\frac{1}{2}- \frac{\mathsf{SINR}_l^m(\mathbf{p}^m)}{(1-\mathsf{SINR}_l^m(\mathbf{p}^m))\sqrt{2\pi \mathsf{SINR}_l^m(\mathbf{p}^m)}}e^{-\mathsf{SINR}_l^m(\mathbf{p}^m)/2}.
\end{array}
\end{equation}

\subsection{Convex Optimization Model}\label{apx:cvxL}
The idea of a {\it convex optimization model} is to treat an exact constrained optimization problem as an individual layer within a deep learning architecture \cite{boydcvxmodel1,boydcvxmodel2,amos2023tutorial}. A general framework can be built as the output of the $k+1$-th layer in a network as the solution to a constrained optimization problem that takes input from the previous layers, and this is given by the following optimization problem \cite{agrawal2019differentiable}:
\begin{align}\label{opt:cvxL0}
\y_{k+1} = \operatorname{argmin}_\y  \;\; & g_{0}(\y ; \bm{\theta}(\y_k)) \nonumber\\
\quad\quad\quad \text { subject to } & g_{i}(\y ; \bm{\theta}(\y_k)) \leq 0, \quad i=1, \ldots, m_{1}, \nonumber\\
& h_{i}(\y ; \bm{\theta}(\y_k))=0, \quad i=1, \ldots, m_{2},
\end{align}
where $\y \in \mathbf{R}^{n}$ is the optimization variable and $\bm{\theta}(\y_k)$ is the parameter vector which depends on the output of the $k$th layer.  

The functions $g_{i}(\y ; \bm{\theta}(\y_k))$ and and $h_{i}(\y ; \bm{\theta}(\y_k))$ are convex and affine in terms of $\y$, respectively. Given a set of parameters $\bm{\theta}(\y_k)$ as input, we can get a solution $\y_{k+1}$ as output. Therefore, we can consider it as a layer that transforms $\bm{\theta}(\y_k)$ into $\y_{k+1}$, which is similar to the conversion layers in deep learning, such as the fully-connected layer, convolutional layer, long short-term memory (LSTM) \cite{goodfellow2016deep}.

\subsection{Explicit Gradients for Backward Propagation}\label{apx:backprop}
To derive the explicit gradients for backward propagation, we have to know $\tilde{\mathbf{p}}$ in terms of $\tilde{\mathbf{r}}$ (both in the logarithmic domain). Rewriting $e^{\mathbf{r}^{m}}-\mathbf{I}=\mathbf{R}^m$, $\text{diag}(\mathbf{G})^{m,-1}=\mathbf{D}^m$ and $\mathbf{G}^m-\text{diag}(\mathbf{G}^m)=\mathbf{F}^m$ for simplicity, we have
\begin{align*}
    \mathbf{p}^{m\star}=(\mathbf{I}-\mathbf{R}^m\mathbf{D}^m\mathbf{F}^m)^{-1}\mathbf{R}^m\mathbf{D}^m\boldsymbol{\sigma}^m.
\end{align*}
To solve for the gradient $\partial\tilde{p}_i^{m\star}/\partial\tilde{r}_j^m$, we first solve for
\begin{align*}
    \frac{\partial\mathbf{e}_i\mathbf{p}^{m\star}}{\partial\mathbf{R}_{jj}^m}=&\ \frac{\partial\mathbf{e}_i^T(\mathbf{I}-\mathbf{R}^m\mathbf{D}^m\mathbf{F}^m)^{-1}\mathbf{R}^m\mathbf{D}^m\boldsymbol{\sigma}^m}{\partial\mathbf{R}_{jj}^m}.
\end{align*}
Recall the formula of inverse matrix w.r.t. a scalar $a$
\begin{align*}
    \frac{\partial\mathbf{A}^{-1}}{\partial a}=-\mathbf{A}^{-1}\frac{\partial\mathbf{A}}{\partial a}\mathbf{A}^{-1}
\end{align*}
and the Woodbury matrix inversion formula
\begin{align*}
    (\mathbf{A}+\mathbf{U}\mathbf{C}\mathbf{V})^{-1}=\mathbf{A}^{-1}-\mathbf{A}^{-1}\mathbf{U}(\mathbf{C}^{-1}+\mathbf{V}\mathbf{A}^{-1}\mathbf{U})\mathbf{V}\mathbf{A}^{-1}.
\end{align*}
With some maths, we can write
\begin{align}\label{eq:explicitgradientx}
    \frac{\partial\mathbf{e}_i^T\mathbf{p}^{m\star}}{\partial\mathbf{R}_{jj}^m}=&\mathbf{e}_i^T(\mathbf{I}-\mathbf{R}^m\mathbf{D}^m\mathbf{F}^m)^{-1}\mathbf{e}_j\mathbf{e}_j^T\cdot\nonumber\\
    &(\mathbf{I}-\mathbf{D}^m\mathbf{F}^m\mathbf{R}^m)^{-1}\mathbf{D}^m\boldsymbol{\sigma}^m.
\end{align}
In~\eqref{eq:explicitgradientx}, we can use the relationship:
\begin{align*}
    \mathbf{R}^{m,-1}(\mathbf{I}-\mathbf{R}^m\mathbf{D}^m\mathbf{F}^m)^{-1}\mathbf{R}^m=(\mathbf{I}-\mathbf{D}^m\mathbf{F}^m\mathbf{R}^m)^{-1}.
\end{align*}
The scalar gradient $\partial\tilde{p}_i^{m\star}/\partial\tilde{r}_j^m$ could be solved by the change-of-variable trick
\begin{align}\label{eq:explicitgradienty}
    \frac{\partial\tilde{p}_i^{m\star}}{\partial\tilde{r}_j^m}=\frac{r_j^m}{p_i^{m\star}}\frac{\partial\mathbf{e}_i^T\mathbf{p}^{m\star}}{\partial\mathbf{R}_{jj}^m}\frac{\partial\mathbf{R}_{jj}^m}{\partial r_j^m}=\frac{r_j^m}{p_i^{m\star}}\frac{\partial\mathbf{e}_i^T\mathbf{p}^{m\star}}{\partial\mathbf{R}_{jj}^m}e^{r_j^m}.
\end{align}
Using ~\eqref{eq:explicitgradientx}, we can apply the explicit formula~\eqref{eq:explicitgradienty} in the backward propagation stage.

\subsection{Mechanisms of the Baselines in Numerical Examples }
In TABLE II in the main text, we examine two categories of baselines:  optimization-based methods and machine-learning methods. 
The optimization-based baselines include the successive convex approximation (SCA) method, as discussed in \cite{pennanen2015decentralized}, and the branch and bound method derived from \cite{zhai2017rate} and \cite{balakrishnan1991branch}. The SCA method stands out for its efficiency, although it may not always lead to globally optimal solutions. Conversely, the Branch and Bound method is noted for its capability to ascertain globally optimal solutions, albeit with a potential compromise in efficiency. 
Due to the limited availability of research studies that specifically focus on the same problem in (7) in this paper,  the aforementioned baselines do not fully align with the problem in (7). However, they do relate to a similar category of wireless optimization problems, particularly the minimization of total power under data rate constraints. Therefore, we adapt these baselines by modifying the methodologies presented in their respective literature to better address the problem in our paper, as detailed below:

\begin{itemize}
    \item SCA method: To address the proposed nonconvex power minimization problem, we draw inspiration from \cite{pennanen2015decentralized} by employing the first-order Taylor expansion. We approximate the concave function $-\sum_{m=1}^M e^{\tilde{r}_l^m}$ at the point $\tilde{r}_l^m(k)$ using the expression $e^{\tilde{r}_l^m(k)} (\tilde{r}_l^m - \tilde{r}_l^m(k)) - \sum_{m=1}^M e^{\tilde{r}_l^m(k)}$. This allows us to construct approximate convex subproblems. By iteratively solving these subproblems, we obtain at least locally optimal solutions.

\item  BnB method: We use Algorithm 1's solution for the upper bound and obtain the lower bound via relaxation of the closest concave envelope for the constraints' exponential functions. Leveraging the BnB mechanism in \cite{balakrishnan1991branch}, we can construct a BnB method to find the global solution for the proposed nonconvex problem. 

\end{itemize}

For machine learning baselines, we refer to the deep neural network (DNN) model in \cite{camana2022deep}, graph neural networks (GNN) in \cite{eisen2020optimal}, and the deep belief network (DBN) in \cite{luo2019deep}. 
\begin{itemize}
    \item DNN: In designing the DNN model, we adhere to the approach presented in \cite{camana2022deep}, incorporating problem parameters such as channel gain, noise power, maximum power, and required transmission rate per user per subcarrier as inputs. The output, an approximate value of $r_l^m$, is utilized to address the convex subproblems; however, we do not treat these subproblems as layers for backpropagation to update the learning parameters.

    \item GNN: For the GNN model, we consider the communication system as a graph $G(V, E)$, with nodes as users and edges as interference. Node features include weights, noise power $n_i$, and maximum power, while edge features represent channel gain $G_{i j}$. We then apply a GNN model to $G(V, E)$ and output the approximate solution, or optimal $p_l^m$. 

    \item DBN: The design of the DBN model is quite similar to the DNN model, but inputs only the system model's channel gain. 
    
\end{itemize}

\bibliographystyle{IEEEtran}
\bibliography{openran}

\end{document}